# Double-side Integration of the Fluorinated Self-Assembling Monolayers for Enhanced Stability of Inverted Perovskite Solar Cells

Ekaterina A. Ilicheva[1,§], Polina K. Sukhorukova[1,2,§], Lev O. Luchnikov[1], Dmitry O. Balakirev[1], Nikita S. Saratovsky[2], Andrei P. Morozov[1], Pavel A. Gostishchev[1], S. Yu. Yurchuk[3], Anton A. Vasilev[3], Sergey S. Kozlov[4], Sergey I. Didenko[3], Svetlana M. Peregudova[2,5], Dmitry S. Muratov[6], Yuriy N. Luponosov[2], and Danila S. Saranin[1]

[1]LASE – Laboratory of Advanced Solar Energy, National University of Science and Technology "MISiS", Leninsky prospect 4, 119049 Moscow, Russia

[2] Enikolopov Institute of Synthetic Polymeric Materials of the Russian Academy of Sciences (ISPM RAS), Profsoyuznaya St. 70, Moscow, 117393, Russia

[3]Department of Semiconductor Electronics and Semiconductor Physics, National University of Science & Technology MISIS, 4 Leninsky Ave., Moscow, 119049, Russia

[4] Laboratory of Solar Photoconverters, Emanuel Institute of Biochemical Physics, Russian Academy of Sciences, 119334 Moscow, Russia

[5]A.N. Nesmeyanov Institute of Organoelement Compounds of the Russian Academy of Sciences, Vavilova St. 28, Moscow 119991, Russia

[6]Department of Chemistry, University of Turin, 10125, Turin, Italy

§: The authors contributed equally to this work

**Corresponding authors:**

Dr. Yu. N. Luponosov luponosov@ispm.ru,

Dr. Danila S. Saranin saranin.ds@misis.ru.

## Abstract

Traps and structural defects at the hole and electron transport interfaces of the microcrystalline absorber limits the efficiency and long-term stability of perovskite solar cells (PSCs) due to accumulation of the ionic clusters, non-radiative recombination and electrochemical corrosion. Surface engineering using self-assembled monolayers (SAM) was considered as an effective strategy for modification of charge-collection junctions. In this work, we demonstrate the first report about complex integration of a SAM for double-side passivation in p-i-n PSCs. Integrating the novel 5-(4-[*bis*(4-fluorophenyl)amino]phenyl)thiophene-2-carboxylic acid (FTPATC) as a fluorinated SAM at the hole-transport interface reduced potential barriers and lattice stresses in the absorber. At the electron-transport side, FTPATC interacted with the A-site cations of the perovskite molecule (Cs, formamidinium), inducing a dipole for defect compensation. Using the passivation approach with fluorinated SAM demonstrated benefits in the gain of the output performance up to 22.2%. The key-advantage of double-side passivation was confirmed by the enhanced stability under continuous light-soaking (1-sun equivalent, 65°C, ISOS-L-2), maintaining 88% of the initial performance over 1680 hours and thermal stabilization under harsh heating at 90°C.

**Introduction**

Halide perovskites (**HP**s) stand out as one of the most promising materials for next-generation photovoltaics (**PV**s)[1], owing unique combination of semiconductor properties and simplified fabrication technology. Solar cells based on HPs are thin-film devices with an absorber of submicron thickness and microcrystalline morphology. HP layers are characterized by strong absorption in the visible range ($10^4$–$10^5$ cm$^{-1}$)[2], suppressed non-radiative recombination dynamics[3], and relatively high charge carrier mobility (up to 10 cm$^2$·V$^{-1}$·s$^{-1}$)[4,5]. Combined with scalable manufacturing methods based on solution-processing (slot-die coating[6], ink-jet printing[7], etc.), this technology shows great potential for reducing capital expenditures (**CAPEX**) at the commercialization stage[8]. To date, perovskite solar cells (**PCS**s) have achieved a power conversion efficiency (**PCE**) record of 26.7%[9], demonstrating benefits over well-developed PV technologies compared to thin-film CdTe and CIGS devices, and offering competitive performance with various crystalline Si cells[9].

However, the use of simplified technological processes complicates the fine-tuning of transport properties and the structural perfection of HP thin-films. Microcrystalline absorber interfaces are characterized with uncompensated defects and a high concentration of structural imperfections[10]. The presence of ionic species and perovskite decomposition products (I$_2$, etc.)[11–13] can induce unfavorable accumulation at the contacts with transport layers or electrodes under an electric field[14,15], potentially triggering electrochemical corrosion. This can result in the decomposition of perovskite and the degradation of the device structure.

The interface engineering using self-assembled monolayers (SAM) was considered as an effective strategy for surface passivation and improvement of energy level alignment in PSCs. The formation of SAM is made possible by the phenomenon of self-assembly, which refers to the spontaneous organization of components into patterns and structures [16]. The formation of SAM is contingent upon the availability of a range of interactions, including covalent bonds, hydrogen bonding, electrostatic interactions, hydrophobic interactions, and steric repulsion between the SAM matrix and the substrate material[17–19]. The formation of covalent bonds in semiconductor devices between SAM and materials employed in electrodes and transport layers (for example, metal oxides) is feasible due to the presence of the so-called anchor group in the SAM. It is most common for acid residues to act as anchor groups, specifically carboxylic and organophosphorus acids. Covalent bonds are formed between the oxygen or hydrogen atoms of the acid group and the metal or oxygen atoms of the semiconductor material[20].

To achieve the optimal properties of SAM, it is possible to vary the anchor group and modify the terminal group of the molecule. A common method for modifying SAM is the introduction of halogen atoms into the terminal group. The incorporation of F, Cl, Br, or I into the molecular composition allows for the manipulation of HOMO and LUMO levels, the work function ($W_F$), the hydrophobicity of the resulting material's surface[21–23], and consequently, the photovoltaic properties of the final device. Of the halogens, the modification of SAM by the introduction of a fluorine atom into the molecule is particularly noteworthy. The fluorine atom is notable for its strong electronegativity, which enables it to form robust hydrogen bonds with the hydrogens of the amino groups present in the organic ligands of the perovskite. This binding action increases the thermal stability of the device[24,25]. Furthermore, the surfaces formed by fluorinated molecules are typically hydrophobic, providing a protective barrier against moisture-induced erosion[26].

In the actual literature, numerous reports highlight the advantages for application of SAM for hole-transport interlayers, mainly in p-i-n device architectures[27,28]. However, surface passivation and stabilization at the absorber/charge transport layer junction on the electron collection side are also required. Double-side passivation, close to c-Si/amorphous-Si heterostructures concept[29], represents a promising approach for stabilization of interfaces and improvement in charge collection efficiency. Several research groups reported the efforts for double side interface modification using organic dielectrics[30], polyelectrolytes[31], halide-containing compounds[32,33] and 2D perovskites[34]. Nevertheless, the stabilization of interfaces and improvements in device efficiency necessitate the development of novel methods and their implementation. The subject of integrating double-sided passivation with modified self-assembled monolayers (SAMs) has yet to undergo rigorous investigation and analysis. In this paper we present a comprehensive study on the use of self-assembled monolayer based on 5-(4-[*bis*(4-fluorophenyl)amino]phenyl)thiophene-2-carboxylic acid (FTPATC) for both-side interface passivation in p-i-n PSC. In our recent research[35], we demonstrated devices based on TPATC material that exhibited excellent performance. To further enhance the material, TPATC underwent a modification process involving the introduction of two fluorine atoms into the benzene rings of the triphenylamine unit, situated in the para-position relative to the nitrogen atom.

We presented a complex investigation into the effect of FTPATC on the surface and optoelectronic properties of the double-cation perovskite $CsFAPbI_{2.93}Cl_{0.07}$. SAM integration resulted in significant Fermi level pinning, changes in energy level alignment, and qualitative improvements in buried interface morphology. We found that FTPATC suppresses trapping and reduces non-radiative recombination, which in turn decreased the dark currents and the non-ideality factor of the PSCs. Passivation with SAM gained the photoelectric performance from ~20% for the bare device to ~21% for p-side modification and ~22%for double-side configuration, respectively. Interface engineering with FTPATC allowed to reach relevant stability of the devices under continuous photo-stress (ISOS-L-2) exceeding 1750h for $T_{80}$ period. The detailed study of surface properties uncovered the interaction of FTPATC with Cs and FA cations, which led to improved thermal stabilization under harsh heating conditions up to 90°C. We deeply analyzed the specific impact of modifying single interfaces and the synergy of combining approaches with the application of fluorinated SAM.

A gathered insights provides a detailed investigation on changes in thin-film/surface properties and physical processes in solar cells.

## Results and discussions

In our work, the surface passivation of charge-transporting interlayers was achieved through the use of FTPATC as a novel SAM. The synthesis of FTPATC is outlined in **fig.1(a)(b)**. The initial stage of the synthesis involved an *N*-arylation reaction between 1-bromo-4-fluorobenzene (**1**) and aniline carried out under Buchwald-Hartwig conditions. This resulted in the formation of a difluorine-substituted triphenylamine derivative (**2**), which was obtained in 91% yield. Subsequently, a non-functional *bis*(4-fluorophenyl)[4-(2-thienyl)phenyl]amine (**5**) was obtained *via* the bromination reaction of compound **2** with *N*-bromosuccinimide (NBS) in 87% yield followed by the Suzuki cross-coupling reaction with 4,4,5,5-tetramethyl-2-(2-thienyl)-1,3,2-dioxaborolane (**4**) in 74% yield. The final stage involved the carboxylation reaction of the *in situ* prepared lithium derivative of the compound **5**, which resulted in the target compound **FTPATC** being obtained in 70% yield. Further details and experimental protocols can be found in the ESI (synthetic procedures section). All of the compounds were isolated and characterized by $^1$H and $^{13}$C NMR spectroscopy (see **ESI, figs. S1–S8**). **FTPATC** was found to be a crystalline material with a melting temperature of 224 °C and a relatively high decomposition temperature (above 255 °C) (see ESI, fig. S9). Moreover, the absorption maximum of **FTPATC** in the thin film occurs at 393 nm, which circumvents the «parasitic» light absorption by the HTL in p-i-n device architectures. The HOMO and LUMO energy levels for **FTPATC** were determined to be –5.42 eV and –2.61 eV, respectively (see **ESI, fig. S10**).

To evaluate the specific properties of the developed SAM, we analyzed the properties of NiO/FTPATC thin-film structures, multilayer samples with microcrystalline perovskite, and p-i-n solar cells. Briefly, the devices with planar, so-called inverted structures were manufactured with the following architecture (**fig.1(c)**): Glass/ITO (anode)–330 nm/NiO (hole-transport layer, p-type, 20 nm)/SAM interlayer/perovskite absorber (intrinsic, 450 nm)/SAM interlayer/C$_{60}$ (electron transport layer, n-type, 40 nm)/bathocuproine (BCP-Ti$_3$C$_2$, hole blocking layer, 8 nm)/Bi–Cu (metal cathode, 10/100 nm). The absorber was Cs$_{0.2}$FA$_{0.8}$PbI$_{2.93}$Cl$_{0.07}$ fabricated accord to our previous investigation[36]. The synthesized FTPATC was integrated into NiO/perovskite and perovskite/C$_{60}$ interfaces via solution processing. To simplify sample identification, we used the following names: "control" for bare configurations, "p-side" for configurations with SAM at the HTL/absorber junction, "n-side" for configurations with SAM at the absorber/ETL junction, and "d-side" (double-side) for modifications on both sides of the perovskite thin-film.

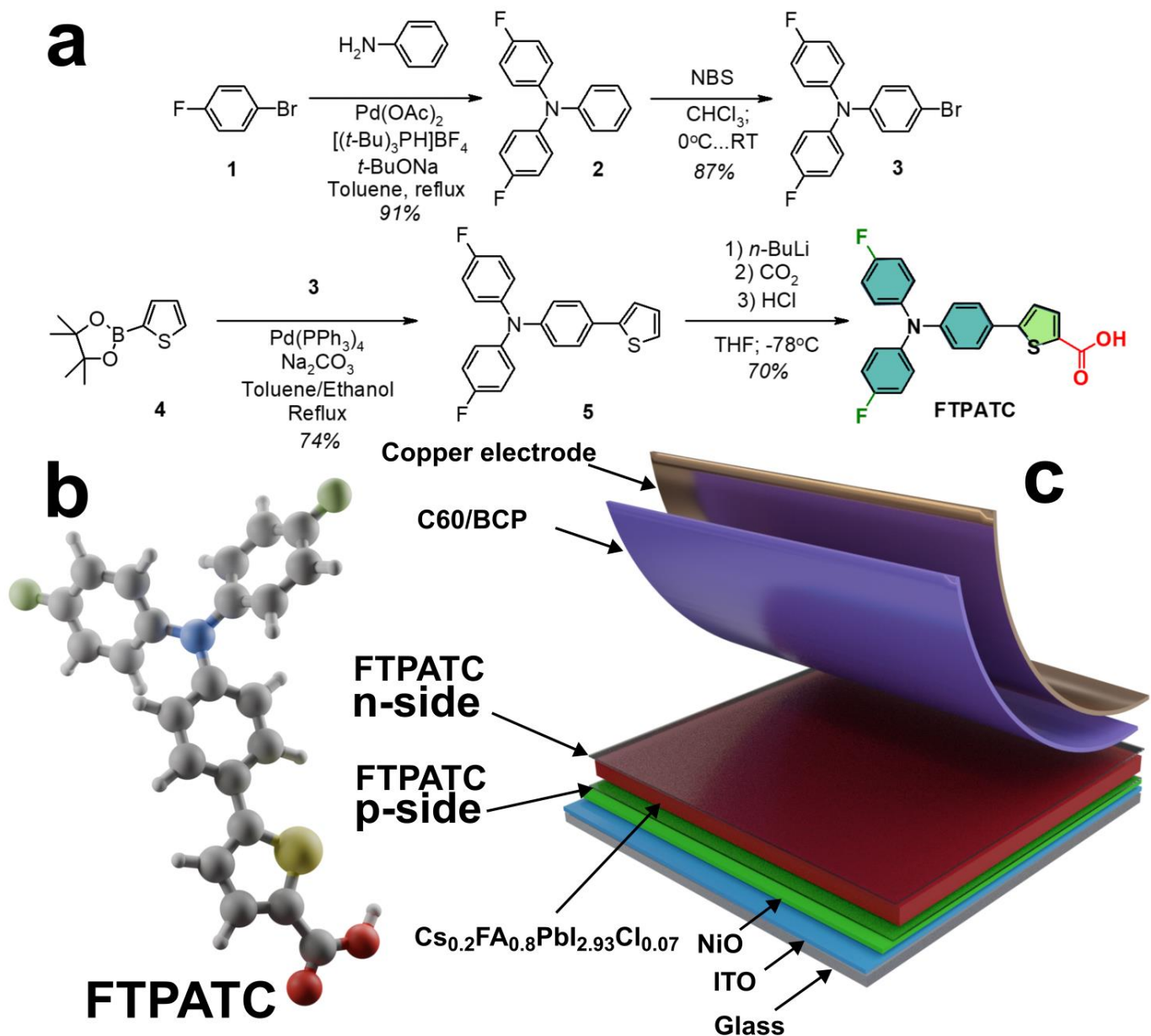

Figure 1 - Synthesis scheme of the compound FTPATC (a); visualization of the FTPATC molecule (b); the device schematics of p-i-n PSCs with integrated SAM interlayers in hole and electron collection interfaces (c)

To estimate the optical properties of FTPATC integrated to HTLs, we measured the transmittance of thin-films, PL spectra and TRPL dynamics for stacks with perovskite absorber and Tauc plot calculations for extraction of band-gap energy ($E_g$) (**fig.S11** in electronic supplementary material (**ESI**)). The obtained transmittance plot (**fig.S11(a)** in **ESI**) indicates FTPATC parasitic absorption in the short-wavelength region (300 - 450 nm), corresponding to its wide band-gap. Analysis of the perovskite absorption spectra with Tauc calculation method (**fig.S11(b)**, **ESI**) showed that band-gap of perovskite thin-film was ~1.58 eV and hadn't relevant changes for p-side, n-side and d-side configurations. The photoluminescence spectra of perovskite absorber with deposited SAM interlayers presented in the **fig.S11(c)** (**ESI**). For multilayer structures with the absorber, an increase in photoluminescence intensity was observed for the p-side and d-side configurations compared to the control sample, whereas no changes in signal amplitude were observed for the n-side configuration. Typically, the quenching of the PL signal in multilayer structures with perovskite thin-films is recognized as a signature of photoinduced charge transfer. Such measurements are one of the routine methods for evaluating photo-induced charge transfer in thin-film HP-based structures, although interpreting these results is not always straightforward. As reported in Campanari's work[37], it is not possible to establish a clear relationship between PL quenching and photoelectric performance. We also observed fluctuations in the peak position for the investigated sample configurations within the range of 1.59 - 1.61 eV, which can be attributed to the reconfiguration of surface trapping states[38]. In our case, the crystallization of perovskite on the FTPATC-modified surface (p-side, d-side) indicates a reduced contribution of non-radiative processes and improved thin-film quality. The obtained trend was supported by TRPL data (**fig.S11(d)** and **tab.S1** in **ESI**). While, the extracted total charge carrier lifetime was almost equal for control and n-side samples (~42 ns), the p-side and d-side showed an increase up to ~50 ns, obviously originated from the presence of TPATC at hole collection interface. Increased carrier lifetime in multilayer samples with HP thin-films is generally indicative of reduced recombination rates, so the charge carriers can exist for longer periods before SRH recombination[39,40].

The specific feature of the SAM for interface integration in PCSs is tuning of the surface wettability. Incorporation of SAM on HTL could enhance hydrophobicity and impacts on crystallization dynamics of the perovskite thin-films. The measurements of the wetting angle (**WA**) on bare and modified NiO surfaces (**fig.S12**, **ESI**) revealed increase of from 26° to 74°. The investigation surface morphology for the control and p-side multilayer stacks (**fig.S13**, **ESI**) exhibited almost equal average grain size ~120 nm.

Despite the absence of notable changes in the morphology of perovskite layers with SAM, the analysis of structural properties could give an insight for differences in lattice parameters, phase composition and possible strain-effects. X-ray phase analysis was conducted on control, p-side, n-side, and d-side samples, as depicted in **Fig.S14** (**ESI**). The predominant phase observed in all samples was the P4/mbm tetragonal phase of β-CsFAPbI$_3$, characterized by peaks at 14.08º, 19.96º, 22.33º, 24.50º, 26.51º, 28.35º, and 31.80º. No additional phases were detected. The introduction of FTPATC notably influenced the peak positions of the perovskite phase (**Fig.2(a)**). Lattice parameters were calculated using the Rietveld method and are presented in **tab.S2** (**ESI**). For the p-side sample, the parameter "a" changed from 8.904 Å in the control sample to 8.923 Å upon FTPATC application. Covering the perovskite with FTPATC resulted in compression of the "c" parameter from 6.314 Å in the control sample to 6.302 Å in both the n-side and d-side samples. In contrast to n-side sample, the d-side had lattice parameters of a = 8.907 Å and c = 6.302 Å. This displays a distinct decrease in crystal unit cell size, leading to the initiation of compressive strain in the perovskite film. As reported [41], compressive strain enhances phase stability of the perovskite film associated with ion migration.

To reveal the features of the interlayer interaction for perovskite/FTPATC, we conducted XPS analysis. The survey spectrum is shown in **Fig. S15** (**ESI**), with changes in elemental composition detailed in **Tab. S2** (**ESI**). The high-resolution F1s spectrum (**Fig. 2(b)**) confirms the presence of fluorine on the back surface of the absorber (electron collection side) as well as at the buried interface with HTL (p-side bottom). Comparing the bare absorber with perovskite/FTPATC (n-side), we observed a 0.15 eV shift in the Cs3d5/2 maximum (**Fig. 2(c)**). Fitting the components of the corresponding Pb4f peaks (**Fig. 2(d)**) indicated area changes in line intensities at 138.9 eV. Notably, for the outer shell electrons of Pb5d5/2 (**Fig. 2(e)**), a shift to lower binding energies (increased kinetic energy) was observed for the n-side configuration. This results in total shift of the maxima valence bend towards Fermi level from 1.55 eV (control) to 1.26 eV (n-side) as shown in the **fig. S16 (ESI)**. N1s spectra presented in **fig. S17 (ESI)**.

In order to elucidate potential chemical interactions for FTPATC with organic cation of the perovskite molecule (formamidinium), the set of liquid-state proton nuclear magnetic resonance ($^1$H-NMR) studies was conducted in DMSO solution (**Fig. 2(f)**). The NMR data indicated the presence of the broad peak, attributed to four equal —$NH_2$ protons, at $\delta$ 8.75 ppm, while the formyl singlet proton peak was observed at $\delta$ 7.84 ppm, for pure formamidinium iodide (FAI), which in accordance with the literature data (full spectra, in **fig. S18, ESI**). The followed addition of FTPATC, the weak carboxylic acid, resulted in a significant change in the spectrum (1:1 molar ratio). The equivalence of the —$NH_2$ protons was no longer observed, and the presence of two signals at $\delta$ 8.99 and $\delta$ 8.64 ppm was noted, while the formyl proton located at $\delta$ 7.84 ppm was recorded as a triplet of triplets. The similar pattern, previously observed in the literature when adding molecular iodine ($I_2$) to FAI in DMSO solution, indicated the formation of strong hydrogen bonds between amino-protons and the triiodide anion ($I_3^-$) via $\eta^3$-allylic complexes shaping, which is not observed for the iodide anion ($I^-$)[42]. Furthermore, the formamidinium triiodide single crystal was isolated and characterized in its individual state[43]. Based on the aforementioned considerations, it can be proposed that a comparable observation was likely to occur for the FAI and FTPATC interaction. The presence of hydrogen bonds ensures the binding of FTPATC on the perovskite surface through the chemical interactions with the organic cation, which, in some studies, is considered as a proxy for better passivation[44]. Slight changes in the cesium peak position, along with NMR data, indicate that FTPATC forms chemical bonds with the A-site cations of the perovskite molecule, specifically Cs and FA. We should note that XPS exhibited considerable changes only in outer shell electrons energy, which could be originated from the locally enhanced electric filed on the surface induced by formation of the dipole[45,46].

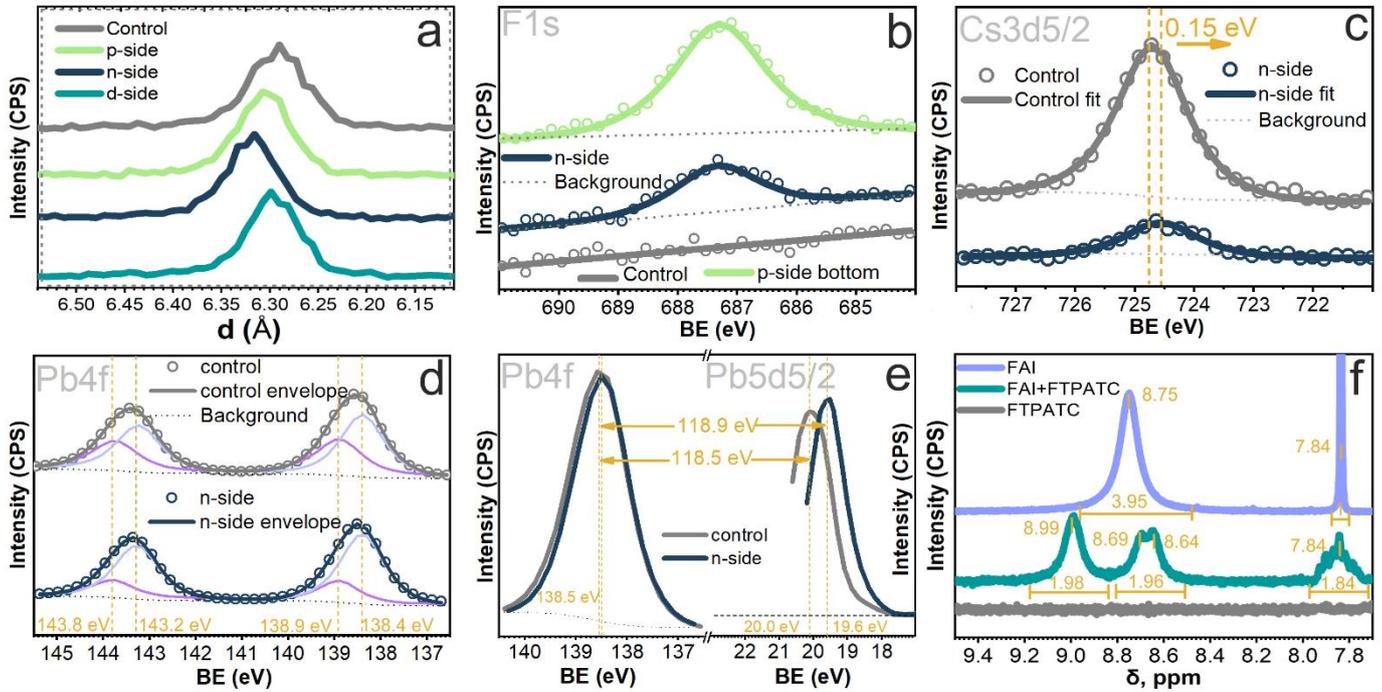

Figure 2 – XRD patterns of perovskite films with FTPATC in different interfaces (a); high resolution XPS spectra of F1s (b); Cs3d5/2 (c); Pb4f (d); Pb5d5/2 (e) for perovskite with and w/o FTPATC and HMR of FAI and FTPATC interaction in DMSO solution.

To explore the possible changes in optoelectronic properties of the modified thin-films, we used KPFM (Kelvin probe force microscope) technique. Analyzing the surface potential of multilayer HTLs with a perovskite layer is not trivial. These studies require consideration of the properties of individual thin-films as well as the buried interface of the absorber, where the surface state plays a crucial role in photo-carrier charge transport. We performed the mapping of the surface potential measurements with calculations of the work-function values for charge transporting layers stacks and de-laminated perovskite absorber with SAM interlayer (**fig.3 (a)** and **fig. S19** in **ESI**). The calculated mean $W_f$ for the surfaces of bare NiO and FTPATC HTL were close, with values of 4.69 and 4.67 eV, respectively. The observed KPFM data showed that the FTPATC interlayer reduces the energy level offset (ΔE) between NiO and the absorber from 0.46 to 0.08 eV. This could potentially reduce the impact of hole accumulation at the interface, reduce the dynamics of non-radiative recombination processes, and increase charge-carrier extraction efficiency. Analysis of the top (electron collection) side of the absorber showed that bare perovskite film has $W_f$ of 4.48 eV, while for n-side configuration we observed a shift to 4.57 eV. Hypothetically, this could affect to energy level alignment with $C_{60}$ ETL which has the position of lowest unoccupied molecular orbit at ~4.2 eV[47]. However, the complex impact of surface modification on quasi-Fermi level splitting under light exposure requires analysis of PSC output.

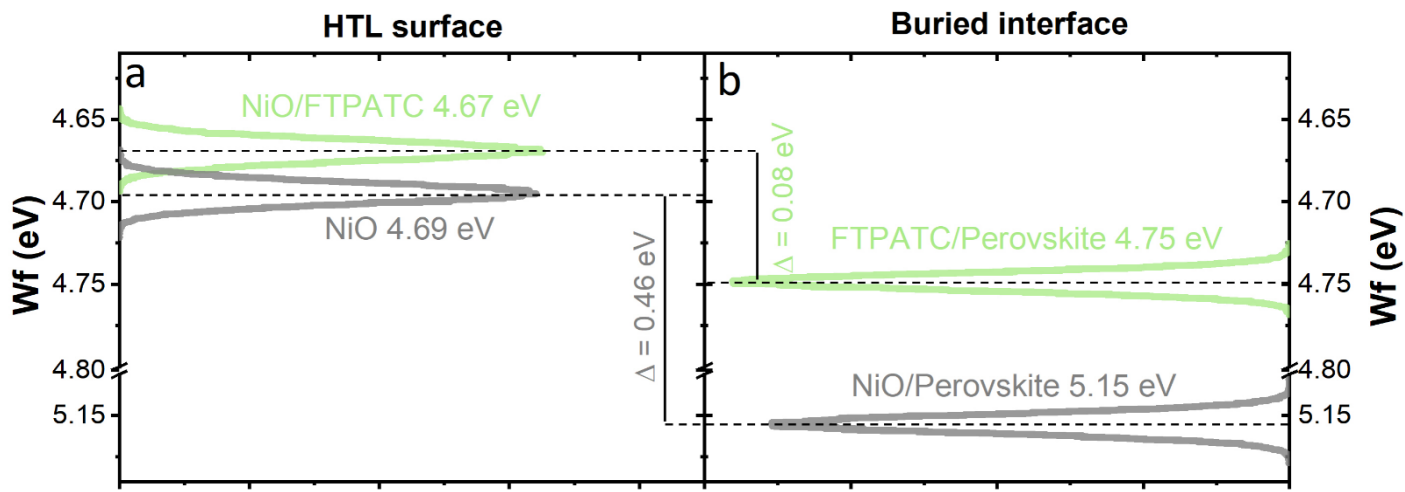

Figure 3 – The $W_f$ values measured for the HTL surface (a) and buried interfaces of the HTL/absorber (b)

Further, we made the analysis of interface passivation with FTPATC on the output of the p-i-n devices. To investigate the photoelectric performance of the fabricated PSCs, we used standard illumination conditions (AM 1.5 G spectra, 100 mW/cm², 25°C) employing AAA Xe solar simulator. The box-charts with statistical distribution of output parameters (open circuit voltage- $V_{oc}$, chort-circuit current density - $J_{sc}$, filling factor– FF, power conversion efficiency - PCE) extracted from the volt-ampere (**IV**) curves presented in the **fig.S20** (**ESI**). The performance of the champion devices is shown on the **fig. 3(a)**. All IV characterization was conducted in the ambient conditions for the encapsulated devices. The best (average) PCE of control PCSs reached the values of 20.2% (18.9%). The devices of p- and d-side configurations showed a relevant increase in PCE up to 22.2% (19.4%) and 21.3% (19.6%), respectively. In contrast, the n-side devices exhibited a slight reduction in PCE with values of 19.9% (18.3%). Primarily, the changes in PCE were driven by $V_{oc}$ and $J_{SC,}$ while the average/best FF values for all device configurations fluctuated in the narrow range of 78-79%. In solar cells, the $V_{oc}$ value related to quasi-Fermi level (**QFL**) splitting, which occurs in the perovskite absorber during photo-injection. Theoretically, $V_{oc}$ and QFLS values should be very close, but experimental results typically demonstrates that $V_{oc}$ is much lower. The large mismatch between QFL and open-circuit voltage in the devices originates from the energy level offset for absorber and charge transporting layers, combined with the processes of the interfacial recombination. This could cause unfavorable bending of the energy levels, leading to a reduction in $V_{oc}$ and affecting charge carrier transport efficiency. Our results highlight that interface modification with FTPATC has a significant impact on $V_{oc}$ changes. The best (average) $V_{oc}$ values for the control devices were 1.08 V (1.07 V). P-side PCSs increased to 1.13 V (1.10 V), while d-side devices exhibited the strongest gain, achieving 1.14 V (1.13 V). Interestingly, for the n-side configuration, we observed a decrease in $V_{oc}$ to 1.06 V (1.03 V), which is lower than the control. Notably, d-side devices, which also have FTPATC on the electron collection side, didn't show this trend. This indicates that reconfiguring energy level alignment and FLP in the absorber should be combined to achieve an optimal $W_f$ shift. The changes in $J_{sc}$ for PCSs with SAM were less dramatic, but statistical analysis clearly showed the effect of interface engineering. The $J_{sc}$ for control PSCs was 23.7 mA/cm² (22.6 mA/cm²). For the p-side configuration, the mean values were similar -22.7 mA/cm², although the champion value reached 24.6 mA/cm². We observed a statistically proved increase in $J_{sc}$ for the n-side configurations at 23.6 mA/cm² (23.1 mA/cm²), as well as for the d-side, which had the highest $J_{sc}$ values at 24.2 mA/cm² (23.6 mA/cm²). The observed results demonstrate that double-sided interface engineering with SAM has a complex impact, with specific effects on both hole and electron-transport interfaces.

Interface modification with FTPATC had a notable effect on device $J_{sc}$ confirmed by external quantum efficiency spectra presented on the **fig.3(b)**. $J_{sc}$ extraction via EQE spectrum integration (AM 1.5 G conditions) demonstrated an increase in photocurrent from 20.8 mA/cm² for the control device to approximately 22 mA/cm² for PSCs with a SAM interlayer. We observed enhanced photoelectric conversion across the entire absorption spectrum of the CsFAPbI$_3$ absorber. Configurations with FTPATC at the HTL interface (p-side, d-side) exhibited a characteristic gain in the short wavelength region of 380 - 420 nm. Photons with higher energy are more efficiently absorbed in the near-surface region of solar cells, indicating improved hole collection efficiency and the absence of parasitic absorption by the FTPATC interlayer.

The stabilization of the output performance was estimated with maximum power point (**P$_{max}$**) tracking (MPPT) of PSCs (**fig.3(c)**) under standard conditions of illumination. Control devices showed absence of the light-soaking effects on $P_{max}$ for 500 seconds. We found that p- and n- side configurations have the opposite trends for the stabilization. P-side devices showed slight increase in $P_{max}$ before saturation (+ ~1.2%), while n-side PCS have a reduction in ~2%. This displays different nature of the traps-filling process under photo-injection, which is typically associated with light-soaking/accumulation effects. Finally, the d-side PSC also had the negligible fluctuation of $P_{max}$, which could be originated from complex impact of energy level alignment.

The transport properties of PSCs were evaluated through an analysis of the dark JV curves (**Fig. 3(d)**). The devices demonstrated relevant diode behavior with strong electrical rectification. The use of SAM effectively reduced the leakage current (**J$_L$**), correlating with the type of interface modification. For the control device, the $J_L$ value (-0.2 V bias) was $10^{-4}$ A/cm². Incorporating FTPATC in single-side configurations reduced this value by an order of magnitude, while for d-side PSCs, $J_L$ reached $10^{-6}$ A/cm². The shunt resistance (**R$_{sh}$**) calculation showed a significant increase, from 3.6 kOhm·cm² for the control to 446 kOhm·cm² for the d-side. The dark volt-ampere characteristic of the solar cell can be divided into four main sections: shunt current (I), recombination current (II), diffusion current (III), and contact resistance (IV). For comprehensive analysis of the charge transport in control and target PSCs, we performed a fitting of dark JVs using 2-diode model (**Fig.S210**, **equations S1-S8**, see **ESI** for the details) and extracted the specific parameters: non-ideality factor (**n**); dark saturation currents (**J$_0$**). The physical meaning of the non-ideality factor is closely related to the dominant recombination processes occurring within the photodiode. For the well-established diode architectures (like Si or A$_{III}$B$_V$ devices), the n values typically are in the range between 1 and 2. When n is close to 1, it indicates that the dominant recombination process occurs in the bulk (neutral) regions of the photodiode (recombination goes primarily due to band-to-band transitions). The increase of n value up to 2 suggests the shift of the dominant recombination process to interfaces involving trap states (so-called Shockley-Read-Hall (SRH) recombination). Due to the presence of two hetero junctions, applying standard models to HP-based p-i-n diode structures is complex. Therefore, the recombination processes in PCSs are not accurately described by an n range of 1 to 2, and corresponding equivalent electrical circuits require modifications. In this work, we used a 2-diode model with series-connected diodes, where the total n for the device was the sum of the individual values. The extracted n (**tab.1**) showed the highest value for the control PSC (2.31), while device with single-side modifications (p- and n- configurations) had reduced and almost equal non-ideality factors ~2.14. D-side PSC demonstrated the synergetic effect of SAM integration, reaching the n=1.58. This trend reveals the beneficial role of FTPATC in the suppression of unfavorable trapping processes at the interfaces. The decrease in $J_0$ for the devices with interface modification (see tab.1) support the observed findings. Calculated series resistance values (**R$_s$**) showed negligible changes with fluctuations of the values in the range of 2-3 Ohm*cm$^2$.

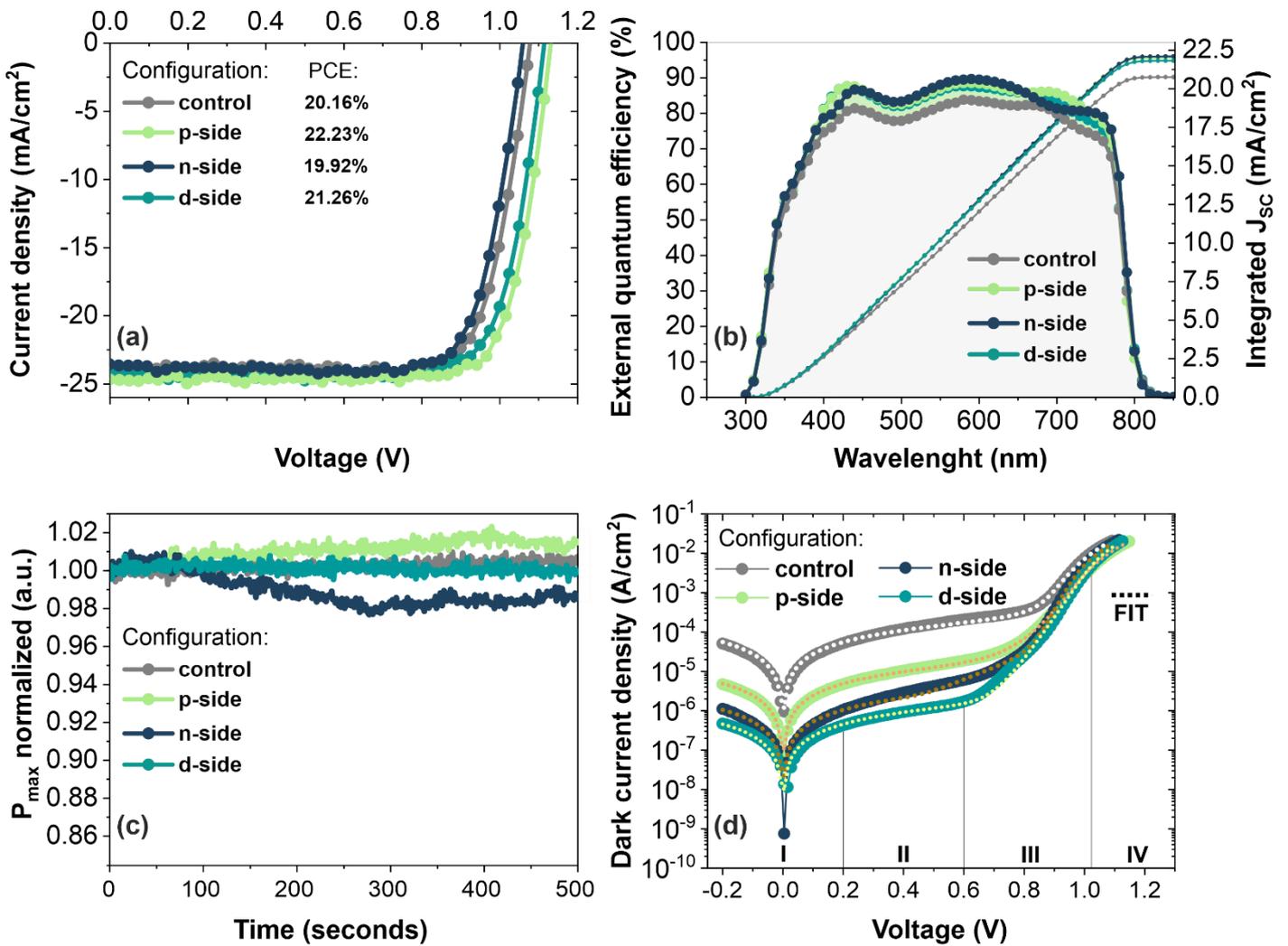

Figure 3 – The JV curves for the PSCs with champion performance (a); external quantum efficiency spectra for the representative PCSs (b); initial stabilization of maximum power under standard illumination conditions (c); dark JV curves with fitting using 2-diode model

Table 1. The extracted parameters for the fitting of dark JV curves using 2-diode model

| Device configuration | $n_1$ | $n_2$ | n ($n_1$+$n_2$) | $J_{01}$, A*cm$^{-2}$ | $J_{02}$, A*cm$^{-2}$ | $R_s$, ohm*cm$^2$ | $R_{sh}$, ohm*cm$^2$ |
|---|---|---|---|---|---|---|---|
| Control | 1.175 | 1.131 | 2.306 | 2.51E-13 | 1.46E-6 | 2.46 | 3.61E+03 |
| p-side | 1.121 | 1.022 | 2.143 | 1.95E-14 | 4.99E-7 | 3.19 | 4.07E+04 |
| n-side | 1.082 | 1.066 | 2.148 | 2.26E-14 | 5.83E-7 | 2.57 | 1.95E+05 |
| d-side | 0.874 | 0.704 | 1.578 | 1.15E-14 | 7.95E-9 | 3.21 | 4.37E+05 |

To further analyze charge carrier extraction properties of PSCs, we conducted transient photocurrent (**TPC**) measurements, presented in **Fig. 4(a)(b)**. When the solar cell is irradiated with a short light pulse, photo-generated charge carriers (photocurrent) are collected at the electrodes. Using a square waveform of the light signal, the device operates in **ON** and **OFF** modes, with the corresponding output at signal saturation. Analyzing the dynamics of the signal rise and decay, as well as the shape of the photo-response, allows to evaluate the contribution of recombination processes and charge accumulation impact. The rising and falling times ($t_r$ and $t_f$) were calculated from the change in photocurrent amplitude from 10% to 90% respectively. For the control sample, the $t_r$ was 10.5 μs, while all configurations with the FTPATC interlayer demonstrated more rapid signal dynamics. Notably, single-side modifications of the interfaces showed variations in the TPC time profile. P-side devices exhibited a $t_r$ of 8.7 μs, and the n-side showed the lowest $t_r$ of 5.5 μs. The d-side configuration showed an averaged value of $t_r$ of 7.4 μs. Analyzing the $t_f$ values showed a similar trend in TPC improvements. It is also notable that the control device clearly exhibited an additional slow component in the decay signal. This feature in the transient response of HP-based optoelectronic devices is typically caused by ion migration at the interfaces (uncompensated organic cations, etc. [48]), strain in the crystal structure and trapping processes [49–51]. The increase in $t_r/t_f$ of the transient response can be described by capacitive effects using various models [52–54]. The reduction in transient time for the devices with FTPATC clearly indicates enhanced charge carrier extraction efficiency, with $t_r/t_f$ reductions ranging from 20% to over 100% compared to the control PCSs. Integrating SAM into the electron-transport junction resulted in the maximum improvement of photo-response dynamics and leveling of the signal's capacitive components.

We performed intensity-dependent performance measurements to analyze the difference photo-response (**Fig. 4(c)(d)**). PSCs were illuminated with white LED sources at intensity levels ($I_0$) from 200 to 10,000 Lux, corresponding to indoor low-light conditions. Under conditions of low concentrations of photo-injected photo-carriers and reduced photocurrent, the contribution of recombination processes and transport efficiency allow for a more precise identification of the specific effects induced by interface modifications in PSCs. The $V_{oc}$ values showed a logarithmic dependence ($V_{oc} \propto \ln$ (photocurrent/dark saturation current)[55], whereas the $J_{sc}$ trends were linear (photocurrent $\propto$ carrier generation rate)[55]. Generally, $V_{oc}$ vs. $I_0$ exhibited changes similar to the mean values measured under standard conditions (AM 1.5 G). At $10^4$ Lux, d-side devices had a $V_{oc}$ of 0.02 V higher compared to control and p-side device. However, the relative difference between control and d-side devices increased to 0.05 V at light intensities of $10^2$ -$10^3$ Lx. This points to an increased losses of charge carrier splitting in mentioned low-light intensity range, due to the complex contribution of shuntы and trapping processes. In the analysis of $J_{sc}$ vs. $I_0$, we observed a different behavior for PSCs' performance. Calculation of the slope coefficient (**k**) of the linear dependence indicated that the n-side and d-side configurations were characterized by an increased steepness, whereas the p-side and control configurations had equal values. The fill factor also revealed differences in intensity-dependent photo-response (**fig.S221, ESI**). The n-side and d-side configurations exhibited reduced FF losses, the control and p-side in contrast, revealed more sensitive behavior at $I_0 \sim 10^2 - 10^3$ Lx. The observed IV pattern under low-light illumination indicates improved electron collection efficiency with FTPATC integration at the absorber/$C_{60}$ contact, and clearly highlights the critical impact of diode properties.

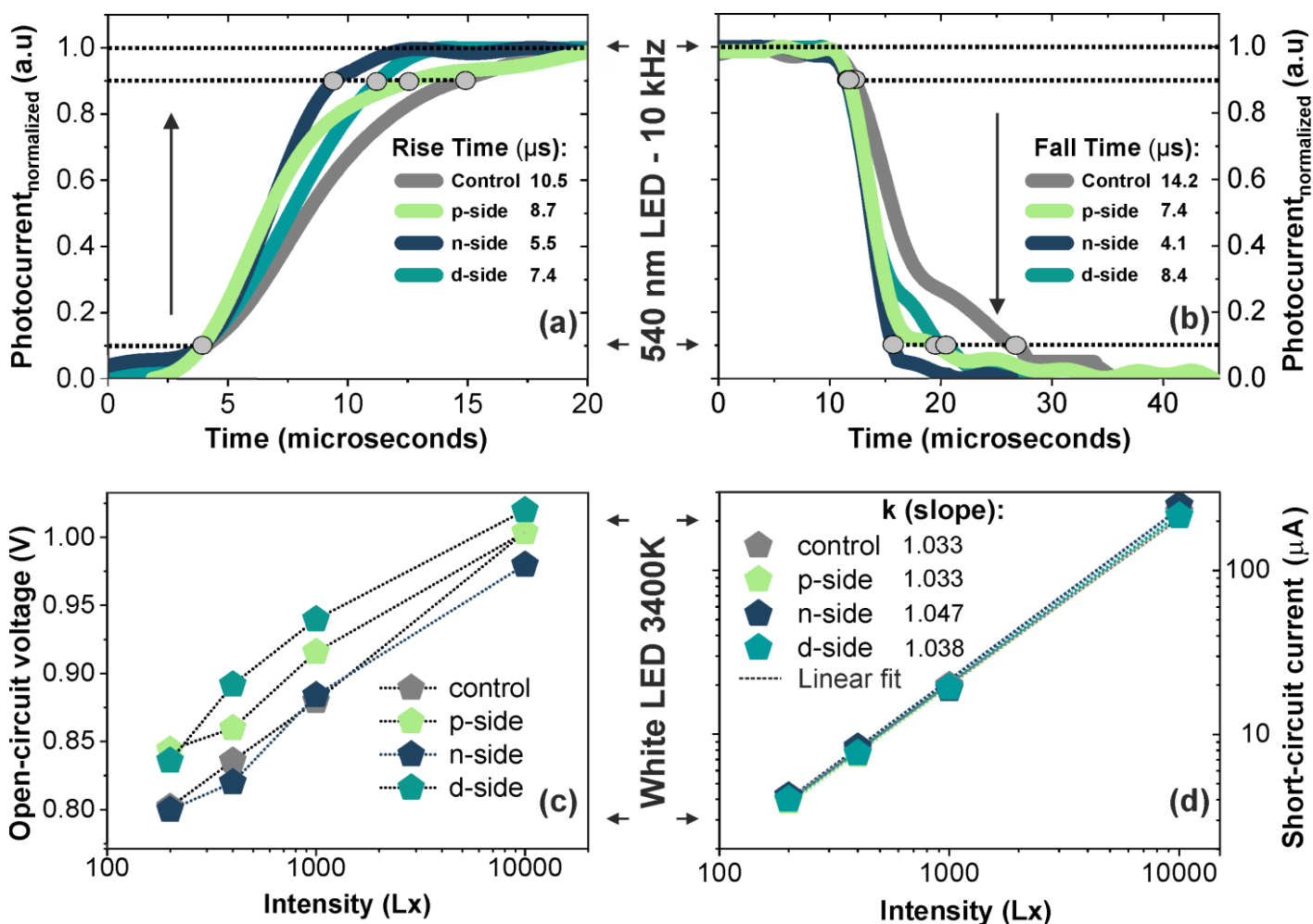

Figure 4 – The transient photo-current measurements in RISE mode (a); transient photo-current measurements in FALL mode (b); $V_{oc}$ vs. light intensity dependence for low-light illumination (c); $J_{sc}$ vs. light intensity dependence for low-light illumination

In parallel, we investigated the stability of the output performance under continuous photo – and thermal stress. Stability tests were conducted on encapsulated devices in ambient conditions with humidity levels not exceeding 60% RH. The conditions of the long-term light-soaking (**LS**, **fig.5**) followed to the ISOS-L-2 protocol[56] (open-circuit mode of operation, 3400K LED illumination, 63.5±1.5°C). All device configuration demonstrated relevant photo-stability over 1600h before $T_{80}$ conditions. Notably, that after soft-heating conditions, PSCs showed an initial drop of the performance that followed by an increase with saturation. Such effects results from the complex processes of the temperature-related changes in charge-splitting and its collection, as well as rearrangement of the traps. Accord to the notes of the ISOS protocol, for the devices with a positive trend in PCE, $T_{80}$ period should be estimated for time at which the efficiency has dropped to 80% of the maxim value, with the complete time from 0 hours to this point quoted as the $T_{80}$. After 1680h of LS, the champion control device reached the level of 81.8%, which was the lowest compared to other configurations. Moreover, statistical distribution of the LS stability data showed that other devices of the batch

(control configuration) have fluctuations in the performance with $T_{80}$ even lower 1000h. P-, n- and d-side devices exhibited slight enhancement of the stability with 90.6, 85.6 and 88.7% of the performance after 1680h supported by narrow data distribution.

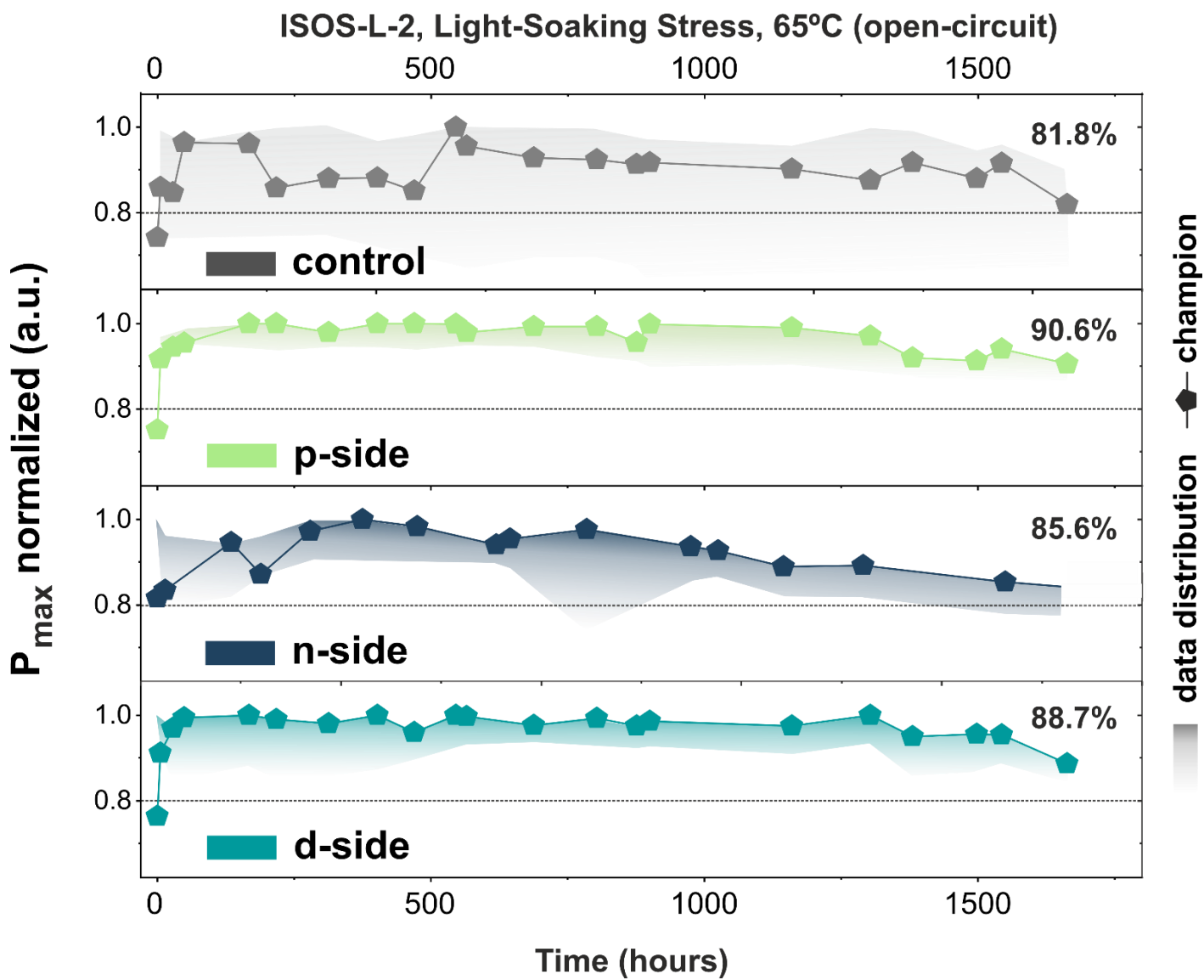

Figure 5 – Stability performance of PSCs under continuous Light-Soaking Stress and soft heating (63.5±1.5°C) for various configurations (a)-(d)

We also conducted studies under harsh heating conditions up to 90°C (ISOS-D-2). The performance loss rate for PSCs was more rapid under enhanced heating stress compared to LS, revealing a thermal stability threshold between 60 and 90°C (**fig.6**). Control PSCs achieved $T_{80}$ after 320 hours, after which their power fluctuated between 80% and 90%. The P-side configuration did not significantly differ from the control sample, with $T_{80}$ at 390 hours. PSCs with FTPATC on the ETL side (n- and d-configurations) showed improved thermal stability up to 600-740 hours, which significantly differed for the trends observed for LS conditions. We found that modification of p-side interface almost doesn't impact the thermal stability at the elevated temperatures (90°C), while n-side modification offers a notable gain in suppression of the power losses. This indicates that corrosion processes are strongly accelerated on the back (electron-collection) side of the p-i-n solar cells harsh heating conditions. Clearly, elevated temperatures accelerate multiple degradation factors in

PSCs. In context of the interface stability topic, the initial corrosion processes related the defect-ion migration[57], as well as release of the gaseous decomposition products from the perovskite absorber[58]. At the elevated temperatures (90- 95°C), the FA-based HPs could undergo the irreversible decomposition[59,60] with formation of $NH_3$, $C_3H_3N_3$, HI, $I_2$ and $PbI_2$. The presence of A-site defects in HP structure (uncompensated ions, vacancies) promotes distortions in lattice environment, and changes in stoichiometry conditions (iodine-rich, etc.), which, in turn, could induce unfavorable phase transition. Moreover, the migration of the organic A-site ions and iodine-containing products towards the back electrodes under electric field impacts to the degradation of metal cathodes[61–63]. The observed results of PSC stabilization with FTPATC indicate a beneficial impact when integrating SAM on the electron collection side. FTPATC interacts with the FA- cation on the surface of the absorber, preventing it from transitioning to a free gaseous state under heating conditions. Additionally, dipole formation at the interface with FTPATC can screen ionic defects during migration in the electric field. Employing double-sided passivation for PSCs combines the advantages of p-side energy level alignment, reduction of lattice stress, and suppression of corrosion processes on the back surface. The developed approach is particularly important for up-scaling issues. Typical exploitation temperature of solar panels in ambient conditions reaches the range of 85-100°C. Therefore, the accelerated corrosion processes at the interfaces should be mitigated not only at standard (25°C) or soft regimes (65°C), but also for harsh heating.

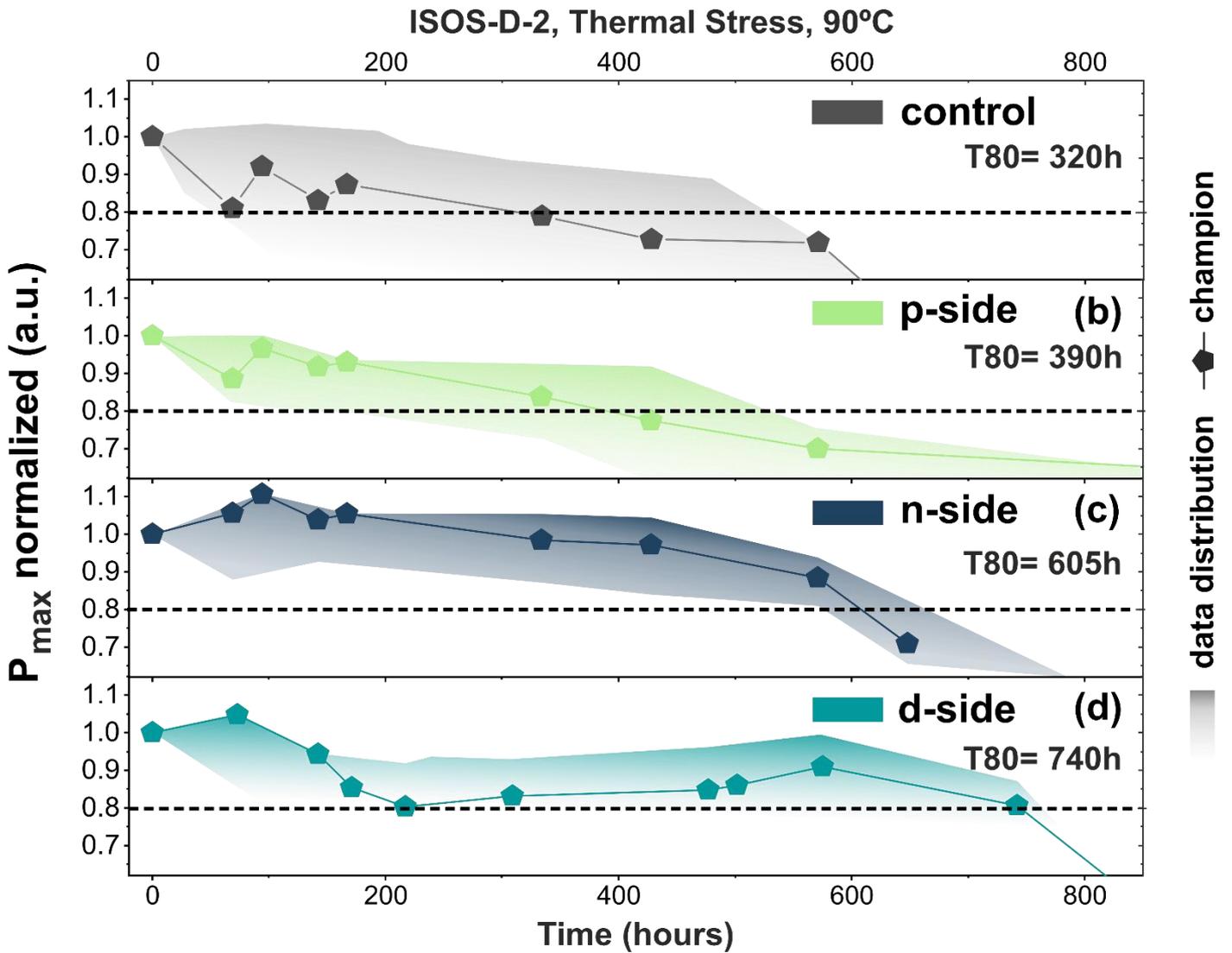

Figure 6 – Stability performance of PSCs under continuous harsh-heating conditions (90°C) for various device configurations (a)-(d)

To the best of our knowledge, the achieved results are well-aligned with state-of-the-art performance in p-i-n solar cells[64]. Additionally, the stability data are superior to alternative double-sided passivation approaches[65–68] or comparable to methods employing more advanced interface engineering[69].

**Conclusions**

In conclusion, our complex investigation introduces an effective approach for double side interface modification in high-performing p-i-n PSCs. The incorporation of FTPATC to charge-collection junctions reduced the potential barriers and improved energy level alignment with perovskite absorber. This promoted suppression of the recombination dynamics at the hole/ electron collection interfaces in the devices and enhanced diode properties. We found that modification of p- and n-type collection junctions has different specifics which impacts to photoelectric output and long-term stability. While p-side integration of FTPATC reduced the strain in the lattice of the perovskite layer, the n-side configuration was characterized by chemical interaction via bonding with A-site cations, which stabilized the interfaces. With this, the double-side passivation strategy with FTPATC has a synergetic effect: p-side modification gains the PCE up to 22.2%, the incorporation of FTPATC to electron-transport boosts the stability of the devices under harsh conditions at the elevated temperatures (90°C). This research offers an important insight for co-modification strategy of the front and back interfaces of the perovskite absorber for balanced optimization of efficiency and stability. Solution-processing with FTPATC could be easily up-scaled for large-area perovskite solar modules in further development.

**CRediT authorship contribution statement**

**Ekaterina A. Ilicheva:** Methodology, Data curation, Writing – original draft, Investigation, Formal analysis, Data curation.

**Polina K. Sukhorukova:** Resources, Data curation, Writing – original draft, Investigation, Formal analysis, Data curation.

**Lev O. Luchnikov:** Methodology, Data curation, Writing – original draft, Investigation, Formal analysis, Data curation.

**Dmitry O. Balakirev:** Methodology, Resources, Data curation, Investigation.

**Nikita S. Saratovsky:** Investigation, Formal analysis

**Andrei P. Morozov:** Methodology, Data curation.

**Pavel A. Gostishchev:** Methodology, Formal analysis.

**S. Yu. Yurchuk:** Methodology, Formal analysis.

**Anton A. Vasilev** Methodology, Formal analysis.

**Sergey S. Kozlov:** Methodology, Formal analysis.

**Sergey I. Didenko** Methodology, Formal analysis.

**Svetlana M. Peregudova**: Investigation.

**Dmitry S. Muratov:** Visualization.

**Yu. N. Luponosov:** Conceptualization, Writing – original draft, Methodology, Funding acquisition, Formal analysis, Data curation, Supervision.

**Danila S. Saranin:** Conceptualization, Writing – original draft, Methodology, Investigation, Funding acquisition, Formal analysis, Data curation, Project administration, Supervision.

**Supporting Information**

Supporting Information, containing synthetic procedures for organic materials (SAM), NMR, mass-spectra, elemental analysis, TGA, CVA, DSC, UV-VIS spectra data, wetting mapping, KPFM data, AMF images, film thickness data, XRD spectra, IV parameters' statistics, TPC plots, IV-temperature parameters, Admittance data, ambient performance of PSMs are available from the Online Library or from the author.

**Declaration of competing interest**

The authors declare that they have no known competing financial interests or personal relationships that could have appeared to influence the work reported in this paper.

**Data availability**

Data will be made available on request.

**Acknowledgements**

The work was supported by the Russian Science Foundation (project № 22-19-00812) - https://rscf.ru/project/22-19-00812/ and conducted at MISiS. NMR spectra were recorded using the equipment of the Collaborative Access Center 'Center for Polymer Research' of ISPM RAS with support from the Ministry of Science and Higher Education of the Russian Federation (topic FFSM-2024-0003).

# Electronic supplementary material for the paper:

# The double-side integration of fluorinated self-assembling monolayers for enhanced stability of inverted perovskite solar cells

# Double-side Integration of the Fluorinated Self-Assembling Monolayers for Enhanced Stability of Inverted Perovskite Solar Cells

**Ekaterina A. Ilicheva**[1,§], **Polina K. Sukhorukova**[1,2,§], **Lev O. Luchnikov**[1], **Dmitry O. Balakirev**[1], **Nikita S. Saratovsky**[2], **Andrei P. Morozov**[1], **Pavel A. Gostishchev**[1], **S. Yu. Yurchuk**[3], **Anton A. Vasilev**[3], **Sergey S. Kozlov**[4], **Sergey I. Didenko**[3], **Svetlana M. Peregudova**[2,5], **Dmitry S. Muratov**[6], **Yuriy N. Luponosov**[2], and **Danila S. Saranin**[1]

[1]LASE – Laboratory of Advanced Solar Energy, National University of Science and Technology "MISiS", Leninsky prospect 4, 119049 Moscow, Russia

[2] Enikolopov Institute of Synthetic Polymeric Materials of the Russian Academy of Sciences (ISPM RAS), Profsoyuznaya St. 70, Moscow, 117393, Russia

[3]Department of Semiconductor Electronics and Semiconductor Physics, National University of Science & Technology MISIS, 4 Leninsky Ave., Moscow, 119049, Russia

[4] Laboratory of Solar Photoconverters, Emanuel Institute of Biochemical Physics, Russian Academy of Sciences, 119334 Moscow, Russia

[5]A.N. Nesmeyanov Institute of Organoelement Compounds of the Russian Academy of Sciences, Vavilova St. 28, Moscow 119991, Russia

[6]Department of Chemistry, University of Turin, 10125, Turin, Italy

## Experimental section

### Substrates

Solar cells were fabricated on $In_2O_3$:$SnO_2$ (ITO) coated glass ($R_{sheet}$<7 Ohm/sq) from Zhuhai Kaivo company (China). Optical measurements were done with use of KU-1 Quartz substrates from LLC Alkor (Russia).

### Materials for synthesis of FTPATC

1-Bromo-4-fluorobenzene, aniline, *N*-bromosuccinimide, palladium(II) acetate (Pd(OAc)$_2$), tri-*tert*-butylphosphonium tetrafluoroborate ([(*t*-Bu)$_3$PH]BF$_4$), sodium *tert*-butoxide (*t*-BuONa), 2.5 M solution of *n*-butyllithium in hexane (*n*-BuLi) and *tetrakis*(triphenylphosphine)palladium (0) (Pd(PPh$_3$)$_4$) (Sigma-Aldrich Co) were used without further purification. Tetrahydrofuran (THF), chloroform, toluene, ethanol and other solvents were purified and dried according to known methods. 4,4,5,5-Tetramethyl-2-(2-thienyl)-1,3,2-dioxaborolane (**4**) was prepared as described elsewhere [S1].

**Materials for perovskite absorber and charge-transporting layers**

NiCl$_2$·6H$_2$O (from ReaktivTorg 99þ% purity) used for HTM fabrication. Formamidinium iodide (FAI, >99.99%) was purchased from Greatcell Solar (Australia), Lead iodide (PbI$_2$, >99.9%) and cesium iodide (CsI, >99.99%) were purchased from Chemsynthesis (Russia) and LLC Lanhit (Russia), respectively. Fullerene-C$_{60}$ (C$_{60}$, >99.5%+) was purchased from MST (Russia). Bathocuproine (BCP, >99.8%) was purchased from Osilla Inc. (UK). Mxenes (Ti$_3$C$_2$) was purchased from Beijing Beike 2D materials (China). The organic solvents: 2-Methoxyethanol (2-ME), dimethylformamide (DMF), *N*-Methyl-2-pyrrolidone (NMP), chlorobenzene (CB), and isopropyl alcohol (IPA) were purchased in anhydrous from Sigma Aldrich and used as received without further purification.

**Synthetic procedures**

**4-Fluoro-*N*-(4-fluorophenyl)-*N*-phenylaniline (2).** Degassed solution of compound **1** (25.000 g, 142.9 mmol), tri-*tert*-butylphosphonium tetrafluoroborate (2.487 g, 8.6 mmol), sodium *tert*-butoxide (17.849 g, 185.7 mmol) and toluene (163 mL) were added to Pd(OAc)$_2$ (0.481 g, 2.1 mmol). The reaction mixture was stirred for 30 min at RT under an inert atmosphere, and then aniline (6.508 g, 38.5 mmol) was added. The reaction mixture was stirred under reflux for 12 h under an inert atmosphere. After completion of the reaction, 250 mL of toluene and 150 mL of distilled water were added to the reaction mixture. The organic phase was separated and washed with water. The solvent was evaporated under vacuum, and the crude product was purified by column chromatography (silica gel; eluent: toluene/hexane = 1/2) to give compound **2** as a yellow powder (18.346 g, 91% yield). $^1$H NMR (250 MHz, DMSO-d6, δ, ppm): 6.88–6.95 (overlapping peaks, 2H); 6.95–7.07 (overlapping peaks, 5H); 7.07–7.18 (overlapping peaks, 4H); 7.20–7.28 (overlapping peaks, 2H). $^{13}$C NMR (75 MHz, DMSO-d6, δ, ppm): 116.10; 116.40; 122.30; 125.80; 129.42; 143.59; 147.43; 156.55; 159.74. Calcd. (%) for C$_{18}$H$_{13}$F$_2$N: C, 76.86; H, 4.66; F, 13.51; N, 4.98. Found C, 77.03; H, 4.71; F, 13.; N, 5.01.

**4-Bromo-*N*,*N*-bis(4-fluorophenyl)aniline (3).** NBS (1.898 g, 10.7 mmol) was added portionwise to a solution of compound **2** (3.000 g, 10.7 mmol) in 50 mL of chloroform. After complete addition, the reaction mass was stirred for 4 hours at boiling. After completion of the reaction, the succinimide precipitate was filtered off and the reaction mass was diluted with 150 mL of chloroform. The organic phase was washed with water, separated, dried over anhydrous Na$_2$SO$_4$ and evaporated under vacuum. After that compound **3** (3.45 g, 87% yield) was obtained as a light yellow powder. $^1$H NMR (250 MHz, DMSO-d6, δ, ppm): 6.75 (d, 2H, *J* = 8.80 Hz); 6.99–7.22 (overlapping peaks, 8H); 7.34 (d, 2H, *J* = 8.99 Hz). $^{13}$C NMR (75 MHz, DMSO-d6, δ, ppm): 113.28; 116.33; 123.20; 126.59; 132.11; 143.02; 146.91; 156.95; 160.15. Calcd. (%) for C$_{18}$H$_{12}$BrF$_2$N: C, 60.02; H, 3.36; Br, 22.18; F, 10.55; N, 3.89. Found C, 60.13; H, 3.39; Br, 22.40; F, 11.02; N, 3.97.

**4-Fluoro-*N*-(4-fluorophenyl)-*N*-[4-(2-thienyl)phenyl]aniline (5).** Degassed solutions of compound **3** (1.500 g, 4.16 mmol) and compound **4** (1.050 g, 5.00 mmol) in a toluene/ethanol mixture (40/6 mL) and

2 M aq. Na$_2$CO$_3$ solution (6 mL) were added to Pd(PPh$_3$)$_4$ (144 mg, 0.125 mmol). The reaction mixture was stirred under reflux for 6 h under an inert atmosphere. After completion of the reaction, 150 mL of toluene and 50 mL of distilled water were added to the reaction mixture. The organic phase was separated and washed with water. The solvent was evaporated under vacuum, and the crude product was purified by column chromatography (silica gel; eluent: toluene/hexane = 1/5) to give product **3** as a light yellow powder (1.35 g, 74% yield). $^1$H NMR (250 MHz, CDCl$_3$, δ, ppm): 6.88–7.14 (overlapping peaks, 11H); 7.18–7.23 (overlapping peaks, 2H); 7.43 (d, 2H, *J* = 8.80 Hz). $^{13}$C NMR (75 MHz, CDCl$_3$, δ, ppm): 116.06; 122.24; 122.52; 124.03; 126.08; 126.82; 127.97; 128.34; 143.57; 144.13; 147.31; 157.38; 160.60. Calcd. (%) for C$_{22}$H$_{15}$F$_2$NS: C, 72.71; H, 4.16; F, 10.45; N, 3.85; S, 8.82. Found C, 72.80; H, 4.21; F, 10.52; N, 3.91; S, 8.94.

**5-{4-[*Bis*(4-fluorophenyl)amino]phenyl}thiophene-2-carboxylic acid (FTPATC).** *n*-BuLi (2.5 M solution in hexane, 1.24 mL, 3.1 mmol) was slowly added dropwise to a solution of compound **5** (1.126 g, 3.1 mmol) in 23 mL of dry THF at –78 °C. After complete addition, the reaction mass was stirred at –78 °C for 1 h. Thereafter, excess dried CO$_2$ gas was slowly bubbled through the stirring reaction mass for 2 h at –70 °C. After the end of the reaction, the cooling bath was removed, and the stirring continued for another 30 minutes with the temperature rising to RT. After completion of the reaction, the mixture was poured in 100 mL of diethyl ether, containing 10 mL of 1 M HCl solution. The organic phase was washed with water (150 mL), then the organic phase was separated, and the solvent was evaporated under vacuum. After purification by recrystallization from ethanol, **FTPATC** (0.88 g, 70%) was obtained as a yellow powder. M.p. = 224 °C. $^1$H NMR (250 MHz, DMSO-d6, δ, ppm): 6.82 (d, 2H, *J* = 8.62 Hz); 7.07–7.24 (overlapping peaks, 8H); 7.32 (d, 1H, *J* = 3.85 Hz); 7.53 (d, 2H, *J* = 8.80 Hz); 7.64 (d, 1H, *J* = 3.85 Hz). $^{13}$C NMR (75 MHz, DMSO-d6, δ, ppm): 116.39; 116.69; 120.92; 123.26; 125.88; 126.92; 132.03; 134.37; 142.85; 148.09; 149.84; 157.12; 160.32; 162.80. Calcd. (%) for C$_{23}$H$_{15}$F$_2$NO$_2$S: C, 67.80; H, 3.71; F, 9.33; N, 3.44; O 7.85; S, 7.87. Found C, 67.93; H, 3.85; F, 9.41; N, 3.52; S, 7.99. MALDI-TOF MS: found m/z 407.2; calcd. for [M]+ 407.1

**Preparation of the Precursors**

Nickel oxide precursor ink was prepared by dissolving the NiCl$_2$·6H$_2$O powder in 2-ME with a concentration of 50 mg/mL. For the perovskite precursor solution, we mixed CsI, CsCl, FAI and PbI$_2$ powders with stoichiometric molar ratio of 0.1:0.1:0.8:1 in DMF:NMP (vol. ratio 6:4) with a concentration of 1.35M, and then stirring at 60 °C for 2 hours. Before coating the solution was filtered with 0.22 μm PTFE filters. For the hole-blocking layer, we used BCP-MXenes solutions prepared by mixing 1 mg of BCP and 1 mg of MXenes powders in 2 mL of IPA. The vial with solution was rinsed in ultrasonic bath for 15 min before stirring at 50 °C overnight to form the dispersion. The FTPATC powder was dissolved in CB at a concentration of 1 mg/ml for NiO/perovskite interface and 0.1 mg/ml for perovskite/C60 interface and the solution was stirred at 50 °C for 1 hour.

**The fabrication of 0.15 cm$^2$ active area PSCs**

The p-i-n-structured solar cell with the following stack ITO/NiO/CsFAPbI$_{3-x}$Cl$_x$/C60/BCP:MXene/Cu. Firstly, the ITO substrates were cut in a dimension of 25×25 cm$^2$ and patterned by UV-laser to isolate 5 mm wide semitransparent ITO electrodes as presented in [https://doi.org/10.1016/j.solmat.2021.111095], then substrates were cleaned with acetone and IPA in an ultrasonic bath and activated under UV-ozone irradiation for 30 min.

The NiO HTL was spin-coated in an air atmosphere with a relative humidity not exceeding 40% with the following ramp: (1 s – 500 rpm, 10 s – 4000 rpm). The deposited NiO layer was annealed at 125 °C for 15 minutes and at 310 °C for 1 hour. For p-side and d-side configuration 1 mg/ml FTPATC solution was spin-coated on top of NiO inside glove box at 3000 rpm (15 s) and annealed at 110 °C (5 min).

$CsFAPbI_{3-x}Cl_x$ film was crystallized on top of HTL stack with a solvent-engineering method. The deposition and crystallization processes of perovskite layers were conducted inside glove box with an inert nitrogen atmosphere. The perovskite precursor was spin-coated with the following ramp: (5 s – 3000 rpm, 30 s – 5000 rpm). 350 μL of EAC were dropped on the substrate on the $22^{th}$ second after starting the second rotation step. Then the substrates were annealed at 90 °C (1 min) and 110 °C (30 min) to form the appropriate perovskite phase.

For n-side and d-side configuration 0.1 mg/ml FTPATC solution was spin-coated on top of perovskite inside glove box at 3000 rpm (15 s) and annealed at 110 °C (5 min). As an ETL 33 nm of C60 was deposited with the thermal evaporation method at $10^{-6}$ Torr vacuum level. The BCP:MXene interlayer was spin-coated at 4000 rpm (30 s) and annealed at 50 °C (1 min) inside glove box. The 100 nm thick copper cathode was also deposited with thermal evaporation through a shadow mask to form a 0.15 $cm^2$ active area for each pixel.

The encapsulation was carried out using a dielectric film made of polyimide, which sealed the active area of the devices, and a protective glass fixed with an epoxy adhesive that hardens under the influence of ultraviolet radiation.

**Characterization**

**NMR spectra.** $^1H$ NMR spectra were recorded using a "Bruker WP-250 SY" spectrometer, working at a frequency of 250 MHz and using $CDCl_3$ (7.25 ppm) or DMSO-d6 (2.50 ppm) signals as the internal standard. $^{13}C$ NMR spectra were recorded using a "Bruker Avance II 300" spectrometer, working at a frequency of 75 MHz. In the case of $^1H$ NMR spectroscopy, the compounds to be analyzed were taken in the form of 1% solutions in $CDCl_3$ or DMSO-d6. In the case of $^{13}C$ NMR spectroscopy, the compounds to be analyzed were taken in the form of 5% solutions in $CDCl_3$ or DMSO-d6. The spectra were then processed on the computer using the "ACD Labs" software.

**Elemental analysis**. Elemental analysis of C, N and H elements was carried out using CHN automatic analyzer "CE 1106" (Italy). The settling titration using $BaCl_2$ was applied to analyze the S element.

**Mass-spectra.** Mass-spectra (MALDI-TOF) were registered on a "Autoflex II Bruker" (resolution FWHM 18000), equipped with a nitrogen laser (work wavelength 337 nm) and time-of-flight mass-detector working in the reflections mode. The accelerating voltage was 20 kV. Samples were applied to a polished stainless-steel substrate. Spectrum was recorded in the positive ion mode. The resulting spectrum was the sum of 300 spectra obtained at different points of the sample. 2,5-Dihydroxybenzoic acid (DHB) (Acros, 99%) and α-cyano-4-hydroxycinnamic acid (HCCA) (Acros, 99%) were used as matrices.

**TGA.** Thermogravimetric analysis (TGA) was carried out in dynamic mode in 30 ÷ 450 °C interval using a "Mettler Toledo TG50" system equipped with M3 microbalance allowing measuring the weight of samples in 0–150 mg range with 1 µg precision. Heating rate was chosen to be 10 °C/min in an argon atmosphere.

**DSC.** Differential scanning calorimetry (DCS) scans were obtained with a "Mettler Toledo DSC30" system with 10 °C/min heating rate in temperature range of 30 ÷ 300 °C in an argon atmosphere.

**UV-Vis spectroscopy.** The absorption spectra were recorded with a "Shimadzu UV-2501PC" (Japan) spectrophotometer in the standard 10 mm photometric quartz cuvette using the THF solution of the **FTPATC** with the concentration of $10^{-5}$ M. All measurements were carried out at RT.

**CV.** Cyclic voltammetry (CV) measurements for **FTPATC** film were carried out with a three-electrode electrochemical cell in an inert atmosphere in an electrolyte solution, containing 0.1 M tetrabutylammonium hexafluorophosphate ($Bu_4NPF_6$) in an acetonitrile and 1,2-dichlorobenzene (4:1) mixture using IPC-Pro M potentiostat. The scan rate was 200 mV s$^{-1}$. The glassy carbon electrode was used as the work electrode. The film was applied to a glassy carbon surface used as a working electrode by rubbing. A platinum plate placed in the cell served as the auxiliary electrode. Potentials were measured relative to a saturated calomel electrode (SCE). The highest occupied molecular orbital (HOMO) and the lowest unoccupied molecular orbital (LUMO) energy levels were calculated using the first formal oxidation and reduction potentials, respectively, obtained from CV experiments in acetonitrile according to following equations: LUMO = e($\varphi_{red}$+4.40) (eV) and HOMO = –e($\varphi_{ox}$+4.40) (eV) [S2, S3].

**Optical measurements.** The absorption spectra of **FTPATC** were recorded with a "Shimadzu UV-2501PC" (Japan) spectrophotometer in the standard 10 mm photometric quartz cuvette using THF solutions of the corresponding compounds with the concentrations of $1 \times 10^{-5}$ mol L$^{-1}$. All measurements were carried out at RT. The optical properties in films were studied using a SE2030-010-DUVN spectrophotometer with a wavelength range of 200–1100 nm. Time resolved photoluminescence measurements was performed with time correlated single photon counter technique (TCSPC) on Zolix OmniFluo-990 spectrofluorometer. The fluorescence was induced with 375 nm picosecond pulsed laser (CNILaser MDL-PS-375). The signal acquisition was conducted until 15000 counts.

*Device characterization.* *JV*-curves were measured in an ambient atmosphere by Keithley 2401 SMU with settling time of $10^{-2}$ s and voltage step of 24 mV. The performance under 1 Sun illumination conditions were measured with ABET Sun3000 solar simulator (1.5 AM G spectrum, 100 mW/cm$^2$). Solar simulator was calibrated to standard conditions with a certified Si cell and an Ophir irradiance meter. The dark *JV*-curves measurements were performed in the dark box.

The EQE spectra were measured using QEX10 solar cell quantum efficiency measurement system (PV Measurements Inc., USA) equipped with xenon arc lamp source and dual grating monochromator. Measurements were performed in DC mode in the 300–850 nm range at 10 nm step. The system was calibrated using the reference NIST traceable Si photodiode. The conformity of spectral response for the measured PSCs was calibrated with Si-solar cell and was compliant to the ASTM E 1021-06 standard. The difference between

Jsc values gathered from IV and those extracted from EQE measurements is mainly related to the different performances of the certified calibration cells used for adjustments to the standard conditions of illumination for the solar simulator and EQE system. The temperature measurements were done using lab-made thermostatic system equipped with 4 thermoelectric modules TEC1-12715 (12W), Arduino control and PID system with real-time temperature control (1°C precision). Maximum power point tracking (MPPT) was performed with the following algorithm (software was developed in LabVIEW): forward J–V scan; calculation of $P_{max}$; $I_{max}$ tracking at $V_{max}$ bias every 1 s and repeat of the cycle every 4 h. An LED projector calibrated to 100 mWcm$^2$ with a certified Si photodiode to simulate 1 Sun conditions was used for illumination during MPPT. Ambient conditions were characterized using calibrated Luxmeter – LT-45 (480-620 nm), insolation meter (500 – 2000 nm) Argus-03 (VNIIOFI), thermohydrometer RGK-30 from UNI-T.

**Admittance Spectroscopy**

To verify the presence of mobile ions, we also performed Admittance Spectroscopy (AS). Notably, the high-power illumination was done through the optical window of the cryostat. Such measurements are particularly sensitive for device active layer properties and its effect of mobile ions. High concentration of mobile charged defects can screen build-in of applied fields on Debye length scale leading to device structure capacitance changes: ($C = \frac{\epsilon\epsilon_0 A}{L_D}$, $L_D = \sqrt{\frac{\epsilon\epsilon_0 k_B T}{q^2 N_i}}$). The same approach used for determining times and frequencies for such defects will follow applied bias. For given diffusion coefficient $D$ time of mobile ion will travel across $L_D$ will be: $L_D^2/D = \tau$. So condition of peak in AS will be: $\omega \cdot \tau = 1 \Rightarrow \tau = \frac{\epsilon\epsilon_0 k_B T}{q^2 N_i D_0} \exp\left(\frac{E_A}{k_B T}\right)$, and then $E_A$ and $D_0$ can be determined from Arrhenius plot.

**KPFM** mappings were carried out with an Ntegra Prima (NT-MDT SI, Russia) commercial scanning probe using a NSG10/Pt (Tipsnano, Tallinn, Estonia) conductive probe with a spring constant of 12 N/m. For Kelvin Probe measurements, the tip scans the surface topography using tapping mode first and then a 1 V AC voltage was applied on the probe near its resonance frequency (~200 kHz) to measure the sample surface potential distribution through a DC voltage feedback loop. The scan rate was set to 0.5 Hz, and a lift scan height of approximately 50 nm was adopted. Tip was calibrated using fresh HOPG surface ($W_f^{HOPG}$ taken as 4.6 eV). Buried interface of perovskite were reviled by pulling out glass substrate glued to the sample .**AFM** for perovskite surface were performed via AIST-NT smart spm1000, using NSG30 (Tipsnano, Tallinn, Estonia) probe in semi contact mode.

**X-Ray diffraction (XRD)** of perovskite layers was investigated with diffractometer Tongda TDM-10 using CuK$_\alpha$ as a source with wavelength 1.5409 Å under 30 kV voltage and a current of 20 mA.

The **X-Ray photo electron spectroscopy (XPS)** measurements were performed using a «PREVAC EA15» electron spectrometer. As a primary radiation source, AlKα (hν = 1486.6 eV, 150 W) were used. The binding energy (BE) scale was pre-calibrated using the positions of Ag3d5/2 (368.3 eV) and Au4f7/2 (84.0 eV) from silver and gold foils, respectively. Perovskite samples films were stored in nitrogen box before measurements.

**Water contact angle** was measured using KRÜSS EasyDrop DSA20.

**NMR spectra**

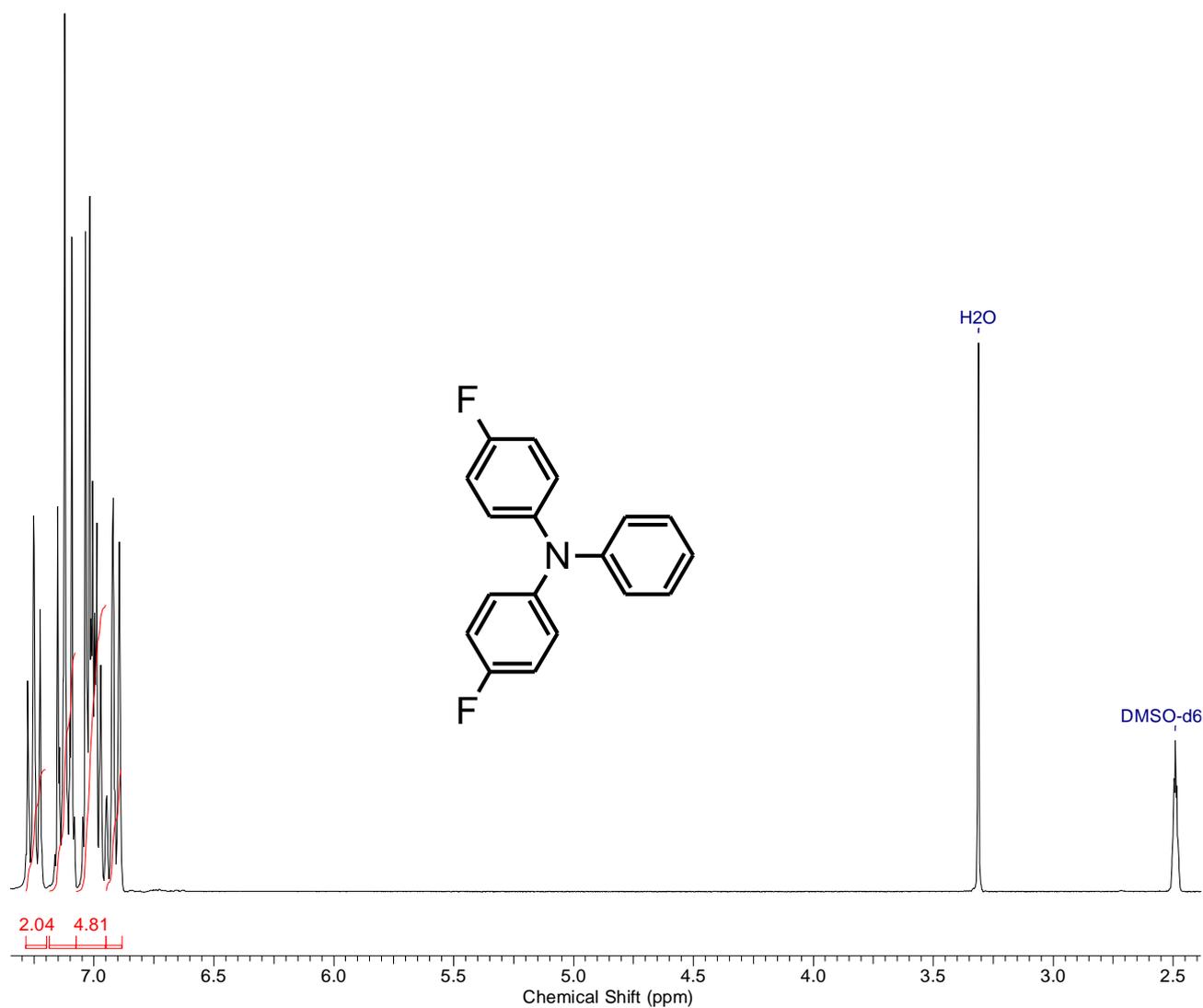

| No. | Annotation | (ppm) |
|---|---|---|
| 1 | DMSO-d6 | 2.49 |
| 2 | H2O | 3.31 |

| No. | (ppm) | Value | Absolute Value |
|---|---|---|---|
| 1 | [6.88 .. 6.95] | 2.081 | 6.46017e+9 |
| 2 | [6.95 .. 7.07] | 4.811 | 1.49334e+10 |
| 3 | [7.07 .. 7.18] | 4.010 | 1.24471e+10 |
| 4 | [7.20 .. 7.28] | 2.045 | 6.34763e+9 |

Figure S1 – $^1$H NMR spectrum of 4-fluoro-*N*-(4-fluorophenyl)-*N*-phenylaniline (compound **2**).

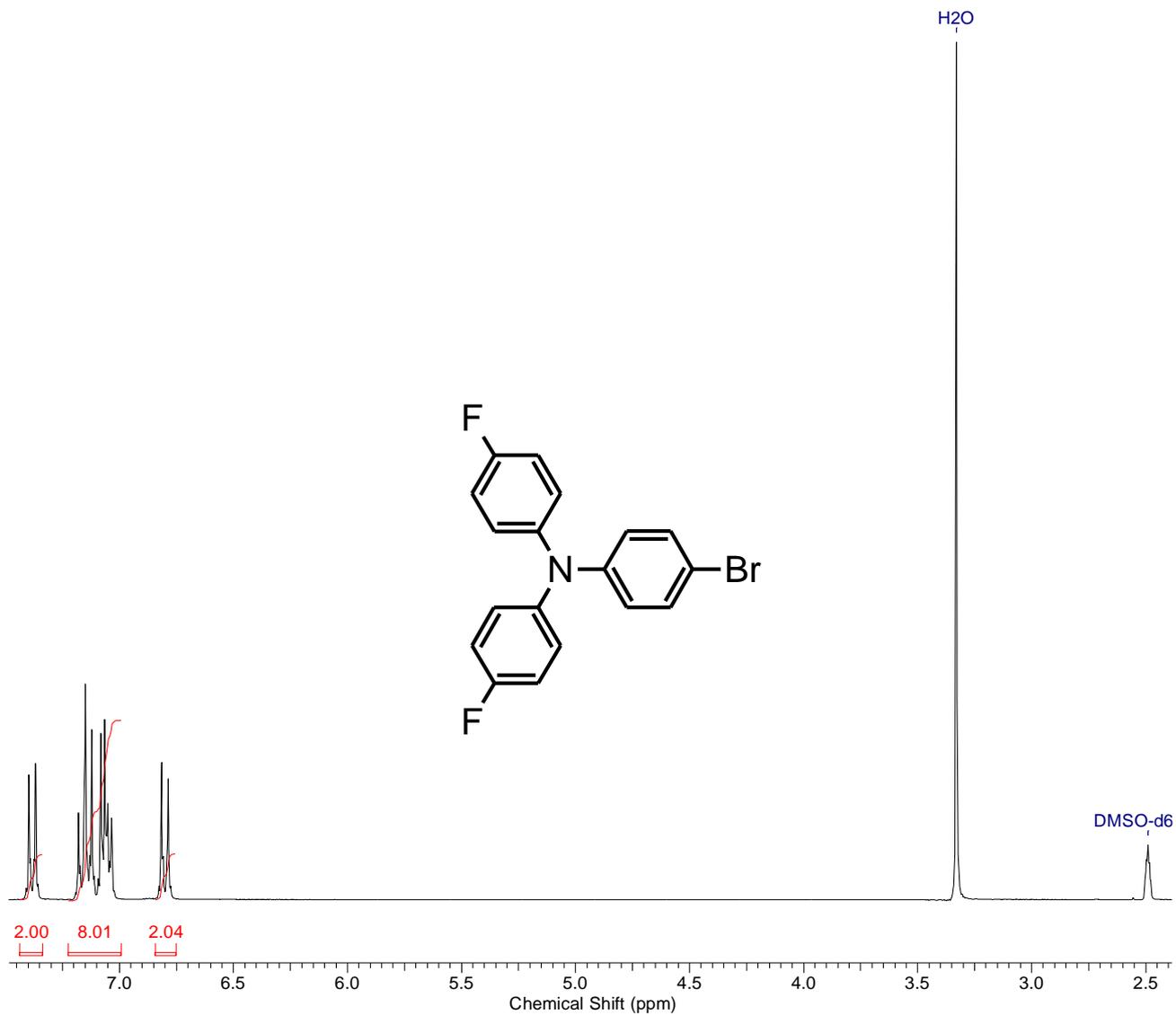

| No. | Annotation | (ppm) |
|---|---|---|
| 1 | DMSO-d6 | 2.49 |
| 2 | H2O | 3.33 |

| No. | (ppm) | Value | Absolute Value |
|---|---|---|---|
| 1 | [6.75 .. 6.84] | 2.039 | 1.73003e+9 |
| 2 | [6.99 .. 7.22] | 8.013 | 6.79811e+9 |
| 3 | [7.34 .. 7.44] | 2.000 | 1.69685e+9 |

Figure S2 – ¹H NMR spectrum of 4-bromo-*N*,*N*-*bis*(4-fluorophenyl)aniline (compound **3**).

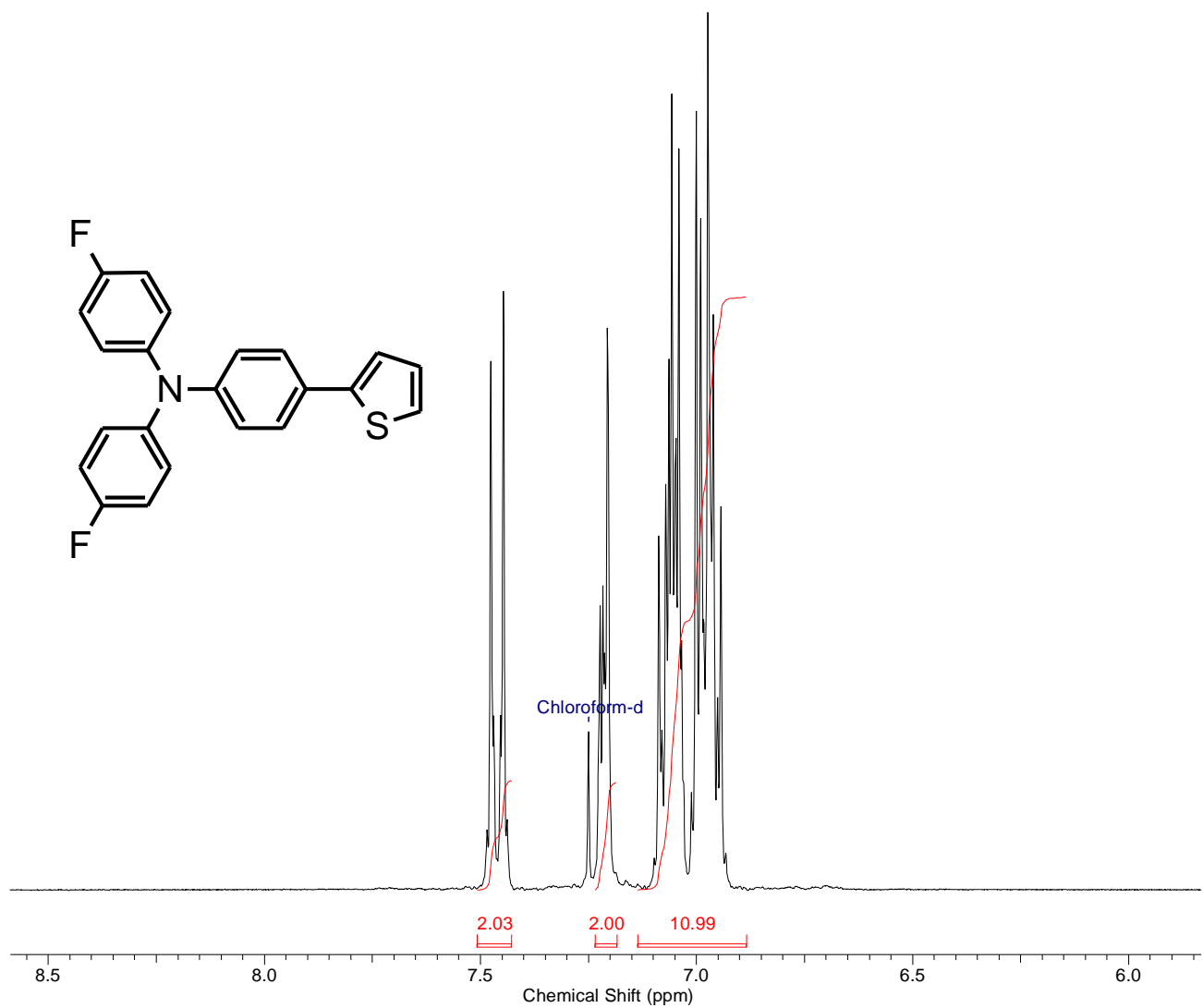

| No. | Annotation  | (ppm) | No. | (ppm)        | Value  | Absolute Value |
|-----|-------------|-------|-----|--------------|--------|----------------|
| 1   | Chloroform-d| 7.25  | 1   | [6.88 .. 7.14]| 10.987 | 3.93506e+10    |
|     |             |       | 2   | [7.18 .. 7.23]| 2.000  | 7.16289e+9     |
|     |             |       | 3   | [7.43 .. 7.51]| 2.027  | 7.25819e+9     |

Figure S3 – ¹H NMR spectrum of 4-fluoro-*N*-(4-fluorophenyl)-*N*-[4-(2-thienyl)phenyl]aniline (compound **5**).

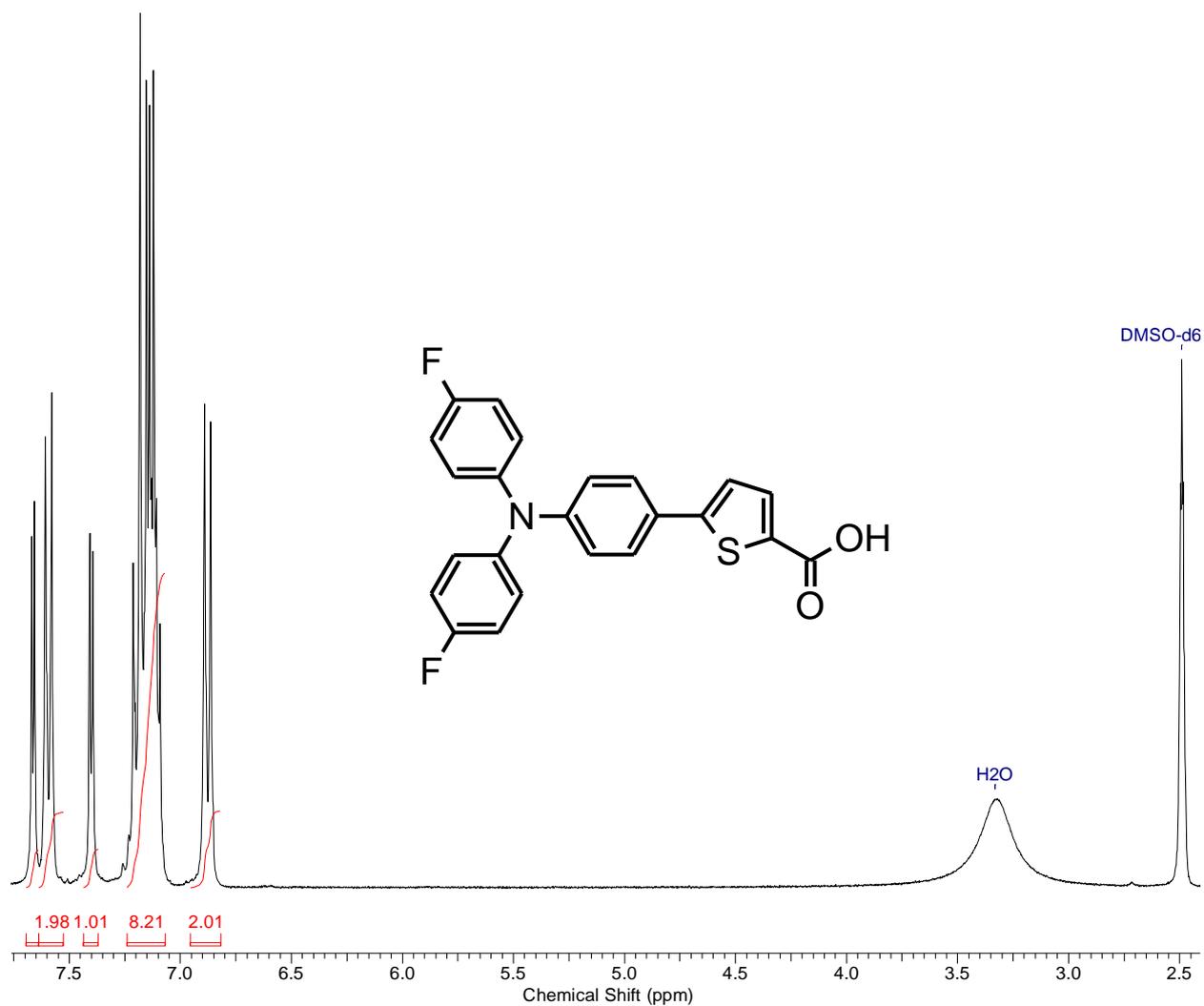

| No. | Annotation | (ppm) |
|---|---|---|
| 1 | DMSO-d6 | 2.49 |
| 2 | H2O | 3.33 |

| No. | (ppm) | Value | Absolute Value |
|---|---|---|---|
| 1 | [6.82 .. 6.95] | 2.008 | 9.02656e+9 |
| 2 | [7.07 .. 7.24] | 8.210 | 3.69065e+10 |
| 3 | [7.37 .. 7.44] | 1.012 | 4.54973e+9 |
| 4 | [7.53 .. 7.64] | 1.978 | 8.89321e+9 |
| 5 | [7.64 .. 7.70] | 1.000 | 4.49538e+9 |

Figure S4 – ¹H NMR spectrum of 5-{4-[*bis*(4-fluorophenyl)amino]phenyl}thiophene-2-carboxylic acid (**FTPATC**).

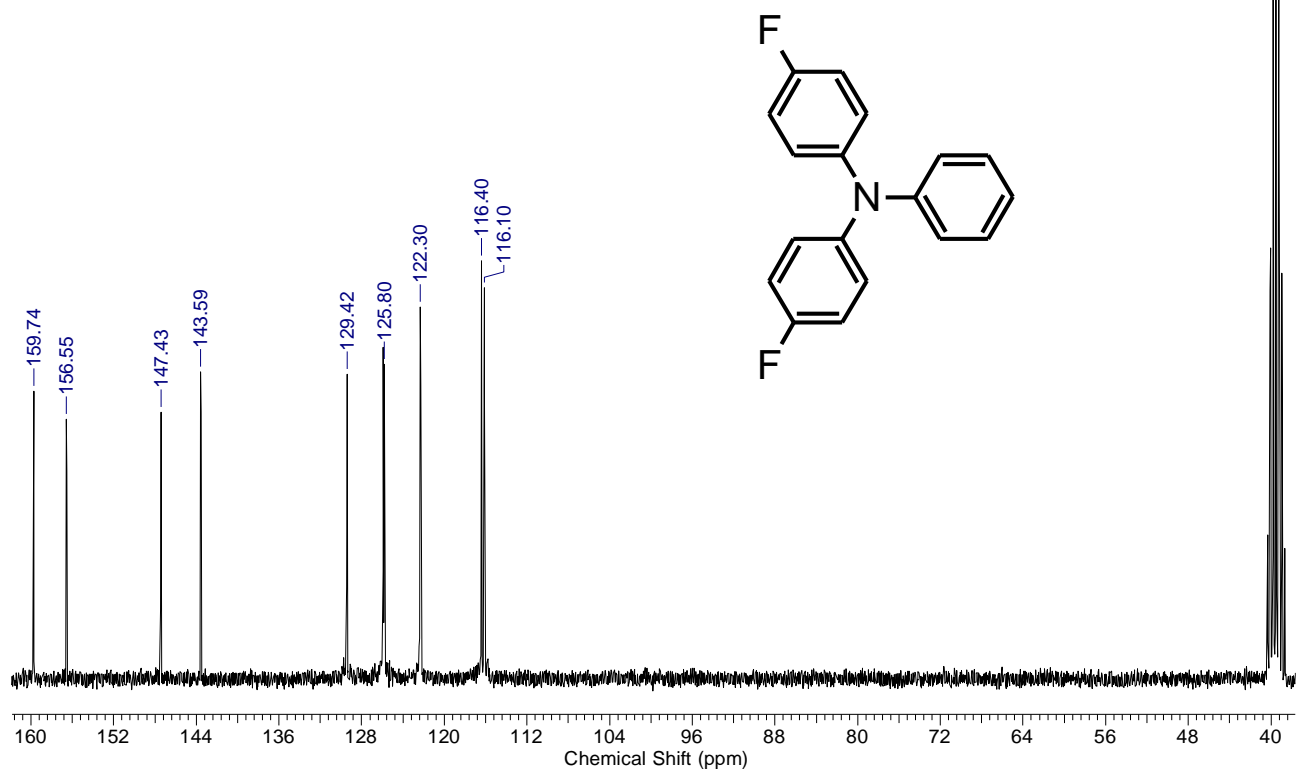

| No. | (ppm) | (Hz) | Height |
|---|---|---|---|
| 1 | 116.10 | 8763.3 | 0.4059 |
| 2 | 116.40 | 8786.0 | 0.4332 |
| 3 | 122.30 | 9231.0 | 0.3866 |
| 4 | 125.80 | 9495.3 | 0.3286 |
| 5 | 129.42 | 9768.4 | 0.3182 |
| 6 | 143.59 | 10837.7 | 0.3210 |
| 7 | 147.43 | 11127.4 | 0.2800 |
| 8 | 156.55 | 11816.3 | 0.2730 |
| 9 | 159.74 | 12056.8 | 0.3011 |

| No. | Annotation | (ppm) |
|---|---|---|
| 1 | DMSO-d6 | 39.50 |

Figure S5 – $^{13}$C NMR spectrum of 4-fluoro-*N*-(4-fluorophenyl)-*N*-phenylaniline (compound **2**).

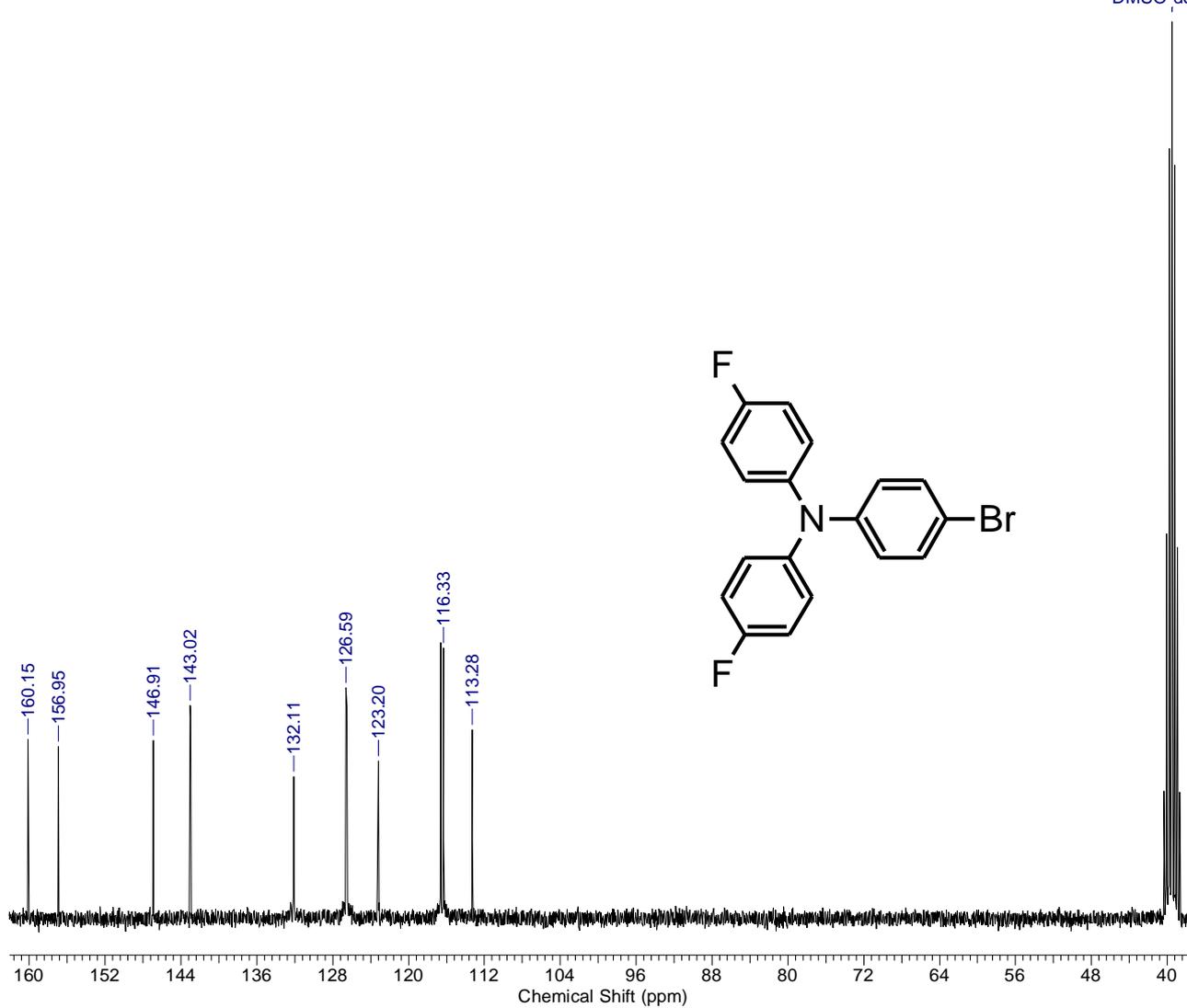

| No. | (ppm) | (Hz) | Height | | No. | Annotation | (ppm) |
|---|---|---|---|---|---|---|---|
| 1 | 113.28 | 8549.9 | 0.2179 | | 1 | DMSO-d6 | 39.50 |
| 2 | 116.33 | 8780.4 | 0.3080 | | | | |
| 3 | 123.20 | 9299.0 | 0.1836 | | | | |
| 4 | 126.59 | 9554.5 | 0.2641 | | | | |
| 5 | 132.11 | 9971.3 | 0.1659 | | | | |
| 6 | 143.02 | 10795.1 | 0.2448 | | | | |
| 7 | 146.91 | 11088.7 | 0.2058 | | | | |
| 8 | 156.95 | 11846.1 | 0.1993 | | | | |
| 9 | 160.15 | 12087.7 | 0.2070 | | | | |

Figure S6 – [13]C NMR spectrum of 4-bromo-*N,N-bis*(4-fluorophenyl)aniline (compound **3**).

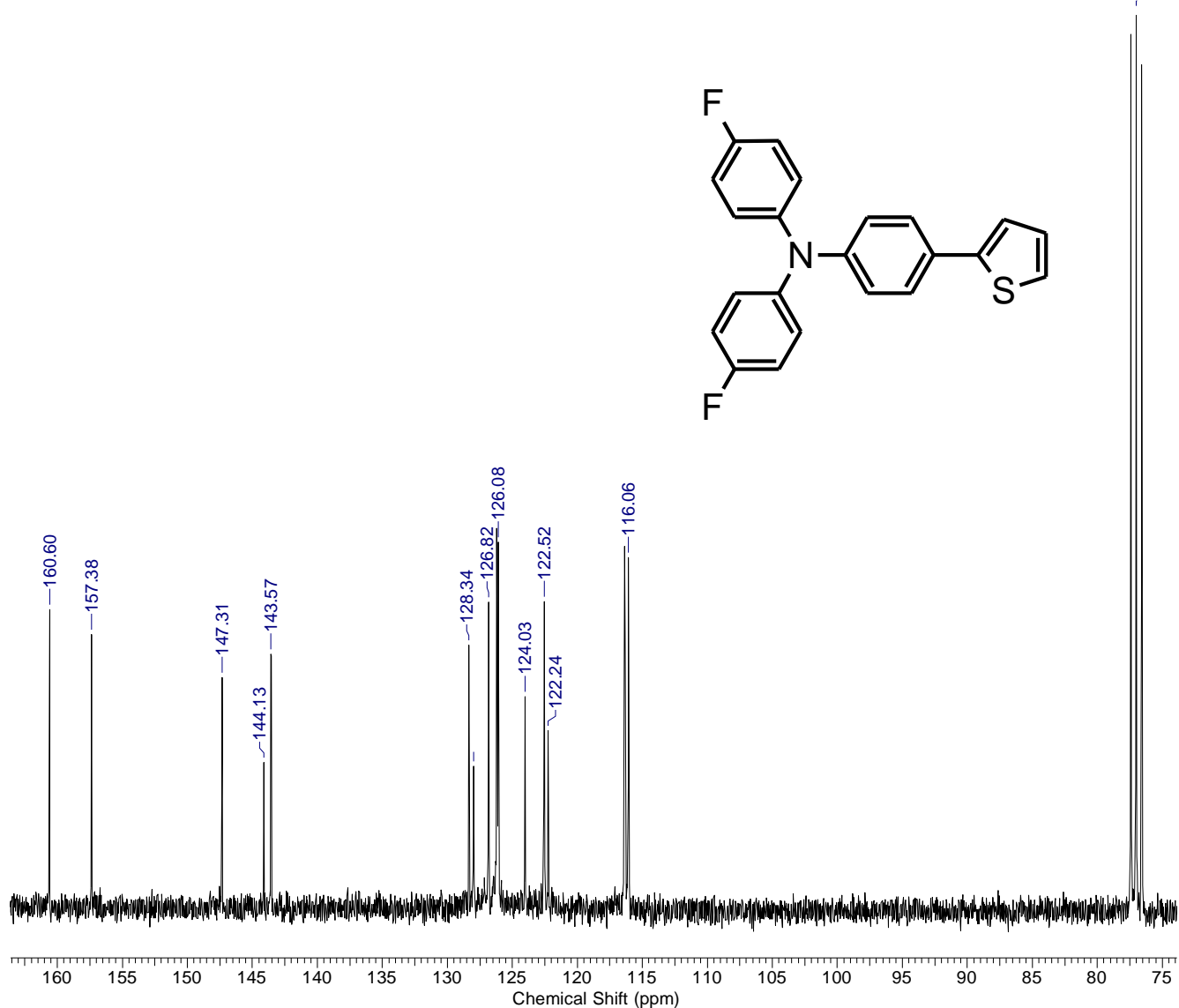

Figure S7 – $^{13}$C NMR spectrum of 4-fluoro-*N*-(4-fluorophenyl)-*N*-[4-(2-thienyl)phenyl]aniline (compound **5**).

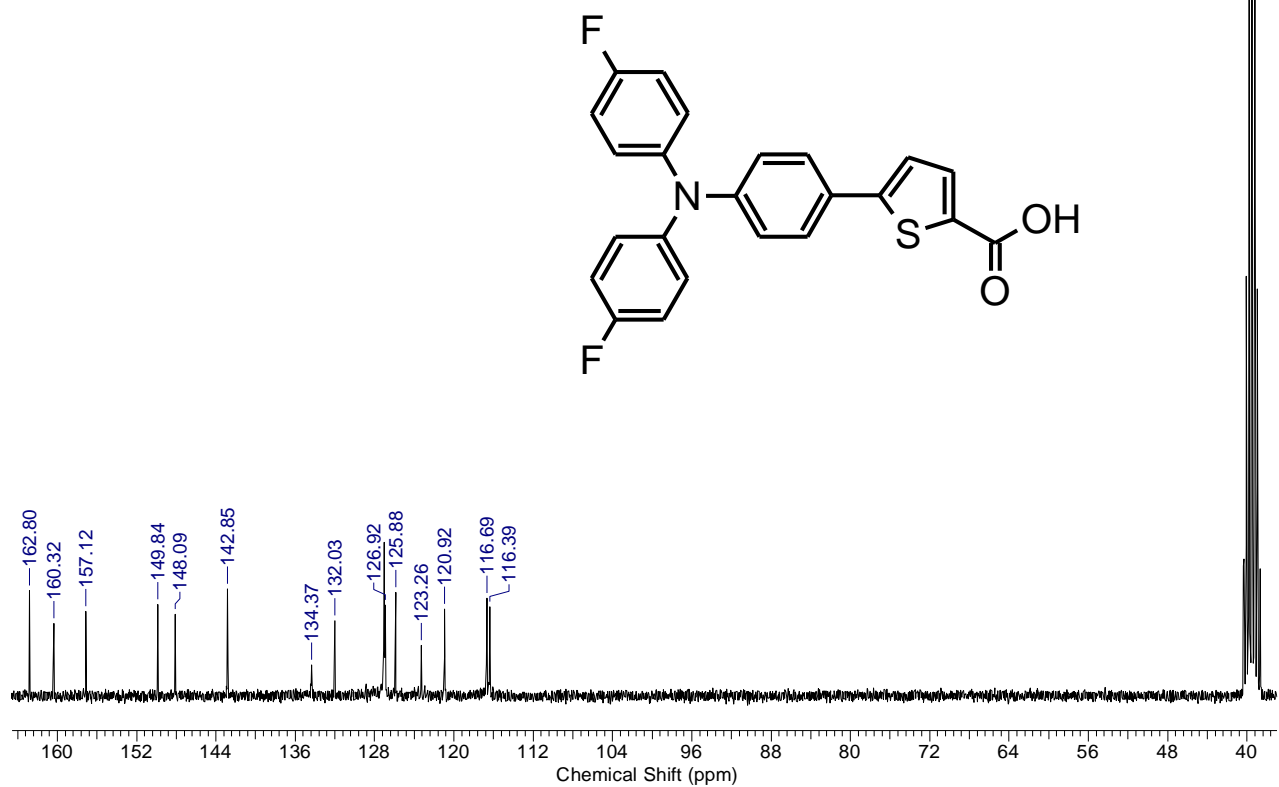

| No. | (ppm) | (Hz) | Height |
|---|---|---|---|
| 1 | 116.39 | 8784.9 | 0.0977 |
| 2 | 116.69 | 8807.5 | 0.1062 |
| 3 | 120.92 | 9126.5 | 0.0950 |
| 4 | 123.26 | 9303.5 | 0.0578 |
| 5 | 125.88 | 9501.4 | 0.1125 |
| 6 | 126.92 | 9579.3 | 0.0989 |
| 7 | 132.03 | 9965.2 | 0.0830 |
| 8 | 134.37 | 10142.2 | 0.0372 |
| 9 | 142.85 | 10781.8 | 0.1157 |
| 10 | 148.09 | 11177.1 | 0.0898 |
| 11 | 149.84 | 11309.8 | 0.1000 |
| 12 | 157.12 | 11858.8 | 0.0927 |
| 13 | 160.32 | 12100.4 | 0.0799 |
| 14 | 162.80 | 12287.9 | 0.1140 |

| No. | Annotation | (ppm) |
|---|---|---|
| 1 | DMSO-d6 | 39.50 |

Figure S8 – $^{13}$C NMR spectrum of 5-{4-[*bis*(4-fluorophenyl)amino]phenyl}thiophene-2-carboxylic acid (**FTPATC**).

## Thermal properties and phase behavior

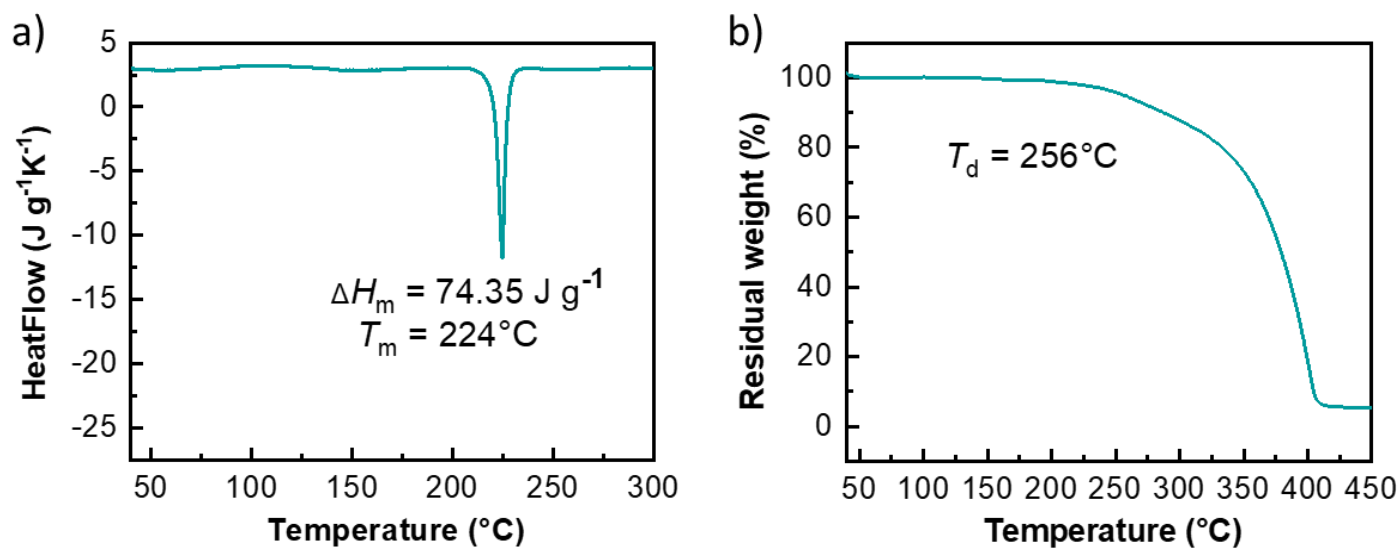

Figure S9 – a) DSC heating scans of **FTPATC** and b) TGA curves of **FTPATC** taken under an inert (argon) atmosphere.

## Optical and electrochemical properties

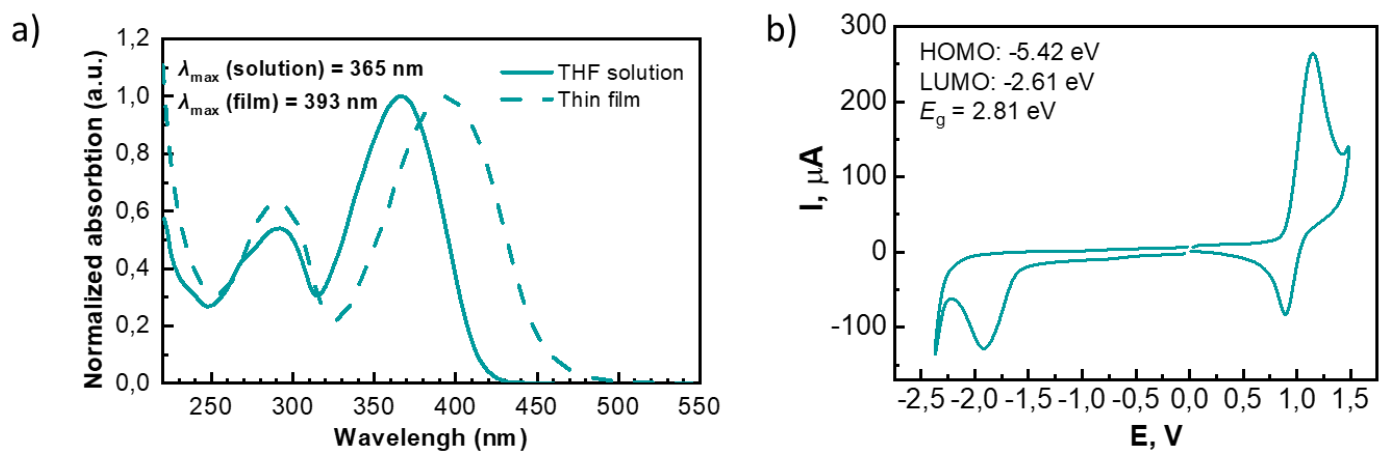

Figure S10 – a) Normalized UV-vis absorption spectra of the **FTPATC** in diluted THF solution and in thin film cast from THF on quartz substrate and b) Cyclic voltammograms for thin film of **FTPATC** in 1,2-dichlorobenzene/acetonitrile (1:4) mixture of solvents, which were recorded with a scan rate of 200 mV s$^{-1}$ using 0.1 M Bu$_4$NPF$_6$ as supporting electrolyte, glassy carbon (s = 2 mm$^2$) as work electrode, platinum plates as counter electrode and SCE (saturated calomel electrode) as reference electrode.

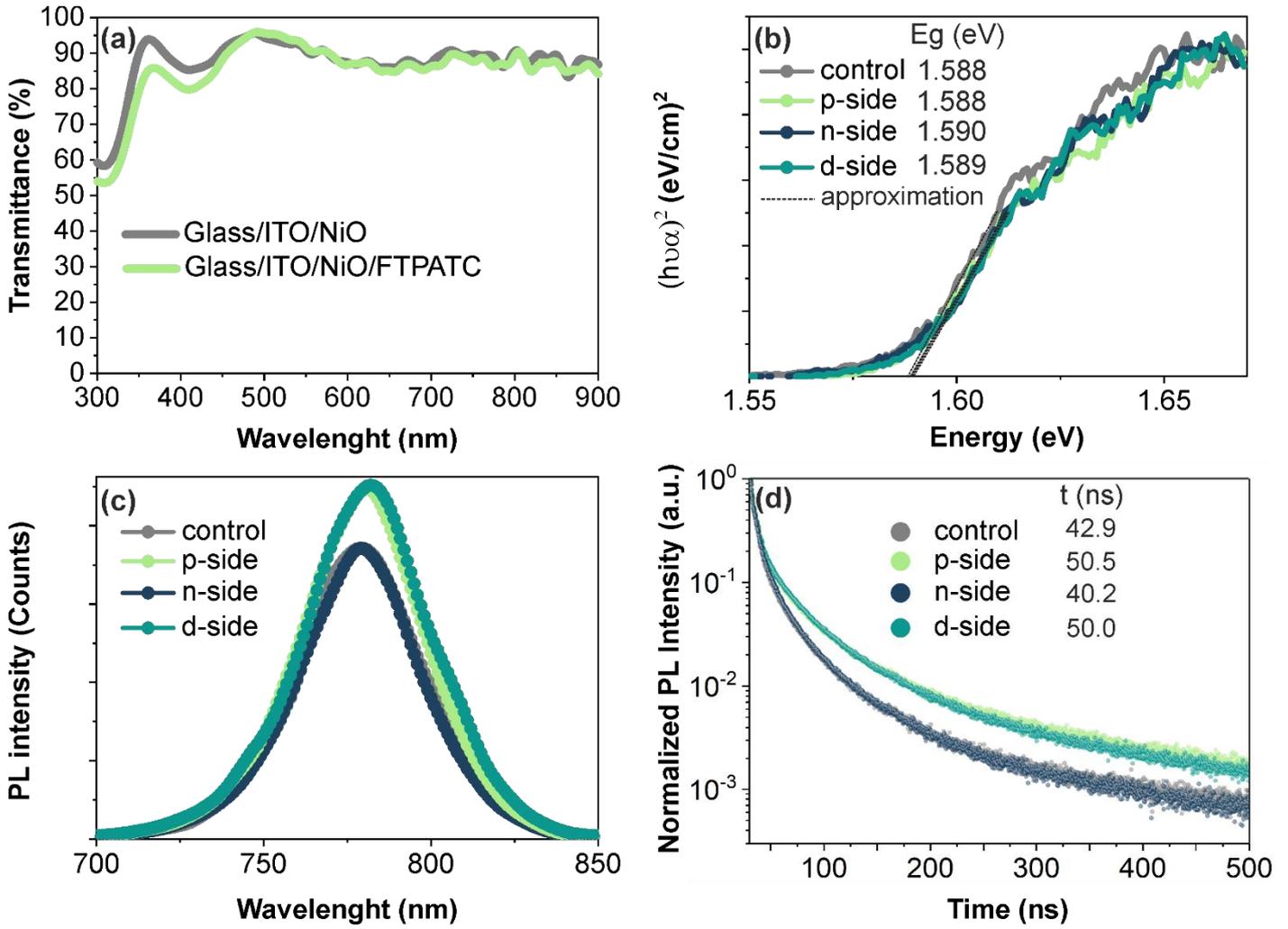

Figure S11 – Transmittance spectra for FTPATC thin-film on HTL/ITO-glass (a) Tauc-plot for multilayer stacks with absorber (b), Photoluminescence spectra of the multilayers stacks with absorber (c), Transient photoluminescence dynamics for the fabricated multilayers stacks with absorber

Table S1 – Analysis of the decay components for measured TRPL signals

| Sample configuration | | $A_1, \div 10^3$ | $\tau_1$ (ns) | $A_2, \div 10^3$ | $\tau_2$ (ns) | $\tau_{average}$ (ns) |
|---|---|---|---|---|---|---|
| **control** | median | 10.0 | 7.5 | 6.8 | 35.5 | 28.9 |
|  | average | 10.2 | 7.6 | 7.0 | 36.6 | 29.9 |
|  | ± std | ± 0.6 | ± 0.5 | ± 0.7 | ± 2.0 | ± 1.8 |
| **n-side** | median | 11.1 | 7.7 | 7.7 | 32.5 | 26.1 |
|  | average | 11.5 | 7.7 | 8.2 | 33.8 | 27.5 |
|  | ± std | ± 1.7 | ± 0.5 | ± 0.9 | ± 2.4 | ± 1.8 |
| **p-side** | median | 10.4 | 8.1 | 10.6 | 42.4 | 37.0 |
|  | average | 10.7 | 8.0 | 10.7 | 42.5 | 37.0 |
|  | ± std | ± 0.5 | ± 0.5 | ± 0.9 | ± 2.1 | ± 1.9 |
| **d-side** | median | 10.5 | 7.9 | 12.1 | 42.2 | 37.4 |
|  | average | 10.4 | 7.8 | 11.4 | 41.1 | 36.2 |
|  | ± std | ± 0.4 | ± 0.5 | ± 2.3 | ± 3.0 | ± 3.4 |

*Notes: $A_1$ and $A_2$ is intensity amplitudes; $\tau_1$ and $\tau_2$ is luminescence lifetimes; $\tau_{average}$ is a weighted average luminescence lifetime; std is standard deviation.*

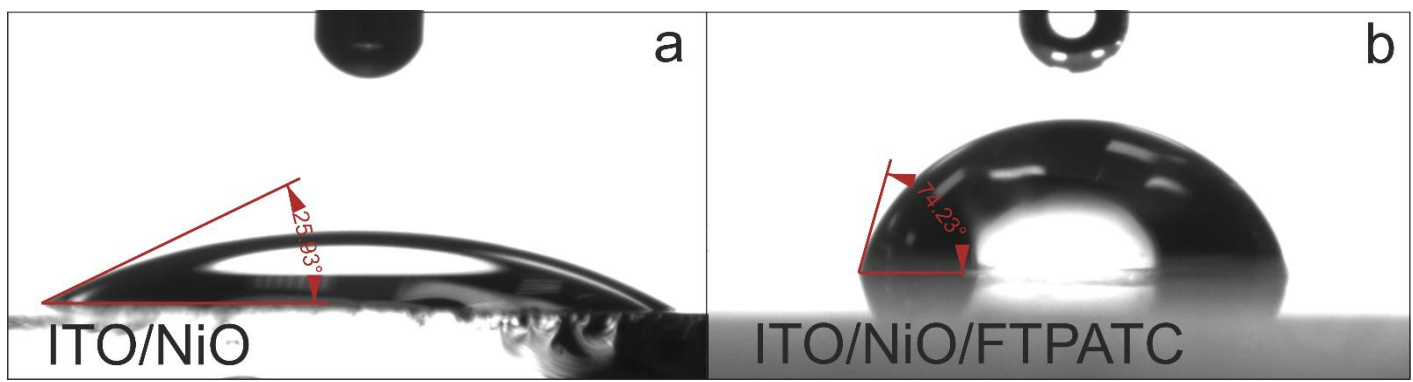

Figure S12 – Wetting contact angle on NiO (a) and NiO/FTPATC (b) films

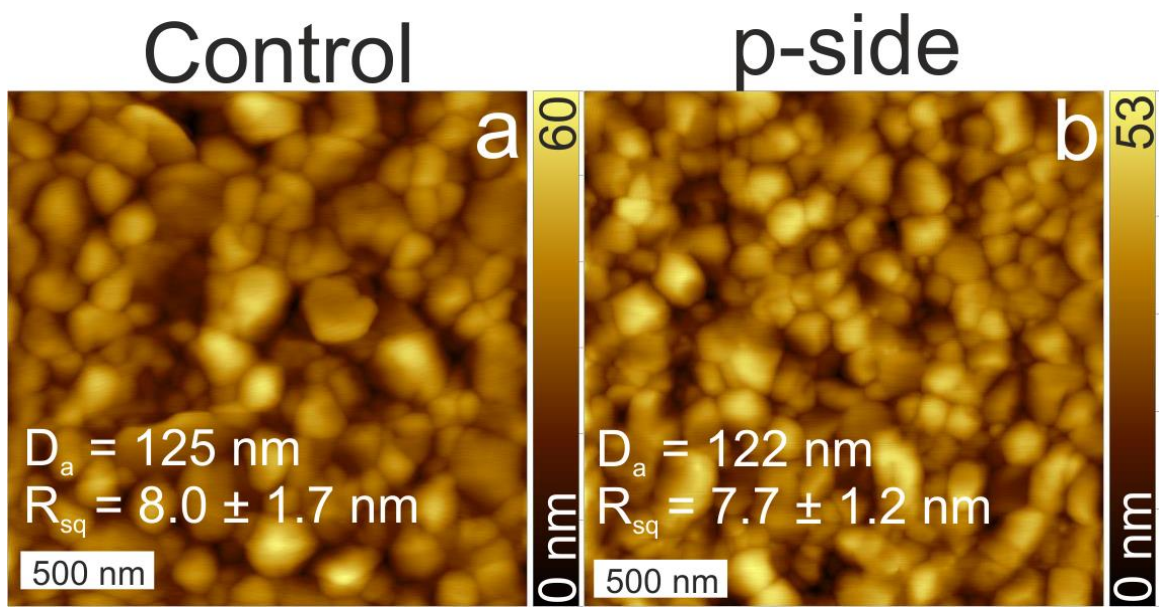

Figure S13 – The AFM images of the bare perovskite absorber (a) and p-side configuration of the sample (b)

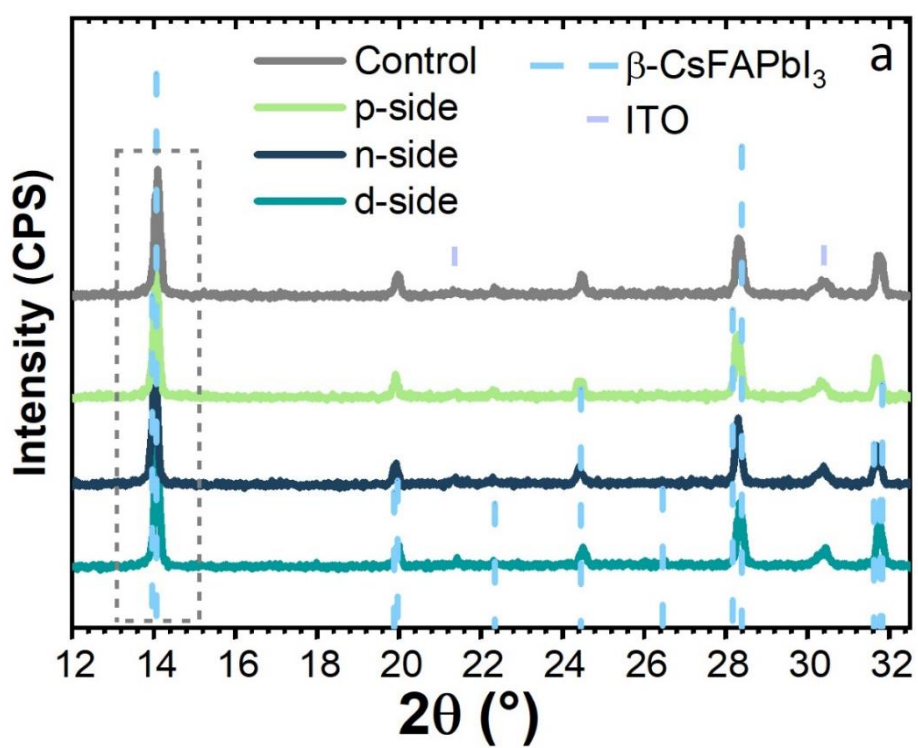

Figure S14 – XRD patterns of perovskite film with FTPATC

Table S2 – lattice parameters of perovskite films with FTPATC

| sample | a (Å) | c (Å) |
|---|---|---|
| control | 8.904 | 6.314 |
| p-side | 8.923 | 6.315 |
| n-side | 8.929 | 6.302 |
| d-side | 8.907 | 6.302 |

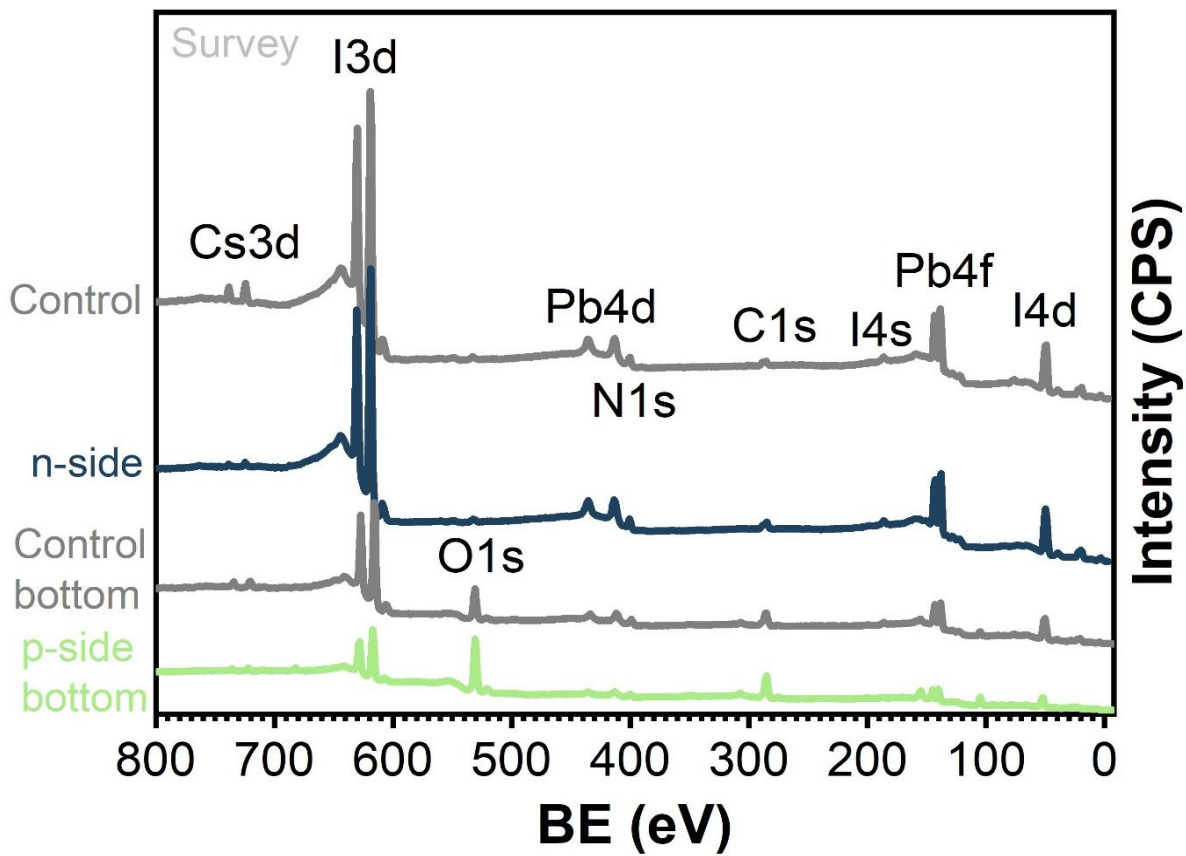

Figure S15 – XPS survey spectra of perovskite films

Table S2 – surface chemical composition of perovskite films (in at%)

|  | C | N | O | F | Cl | I | Pb | Cs |
|---|---|---|---|---|---|---|---|---|
| R.S.F. | 1 | 1.8 | 2.93 | 4.43 | 2.285 | 33.64 | 22.74 | 40.22 |
| control | 25.03 | 14.68 | 4.88 | 0.00 | 0.59 | 40.03 | 13.41 | 1.8 |
| n-side | 34.12 | 13.53 | 4.51 | 0.51 | 1.15 | 32.94 | 12.92 | 0.34 |
| control bottom | 45.86 | 7.10 | 27.94 | 0.00 | 0.77 | 13.19 | 4.59 | 0.43 |
| p-side bottom | 51.2 | 3.25 | 36.62 | 1.68 | 0.86 | 4.54 | 1.64 | 0.2 |

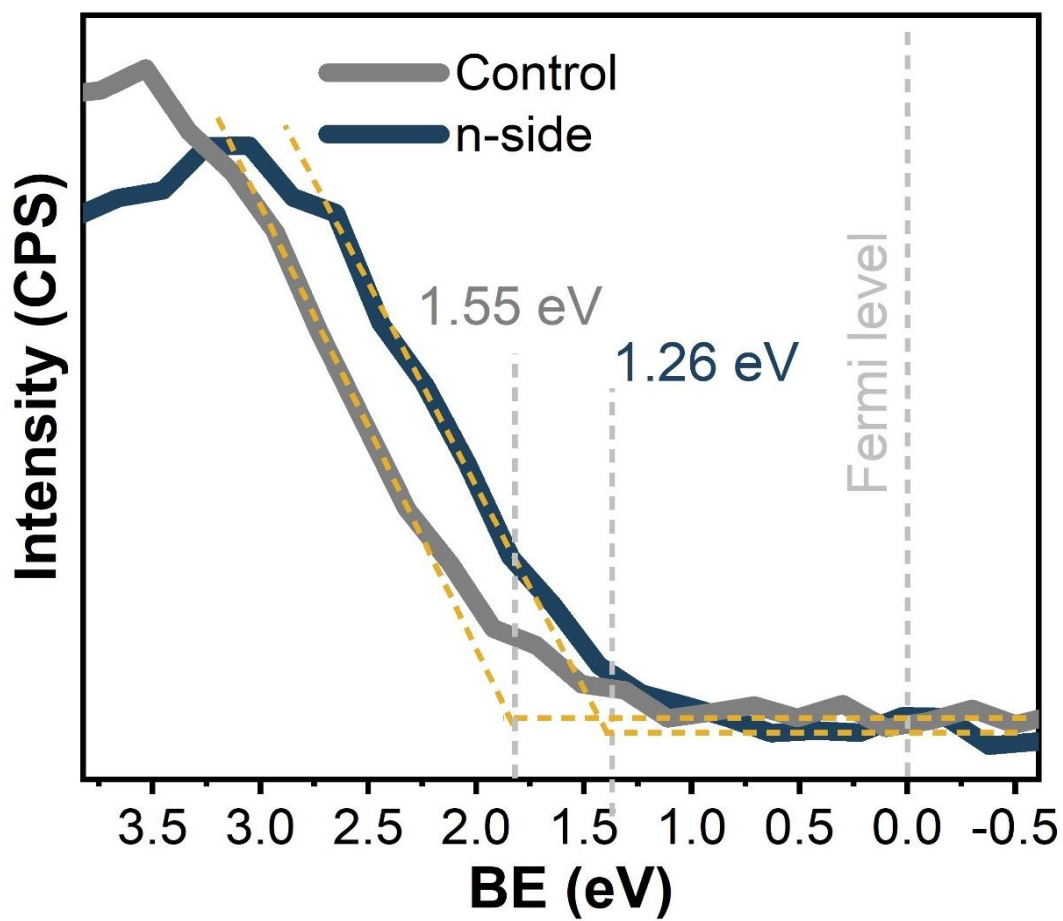

Figure S16 – valence band spectra of perovskite films

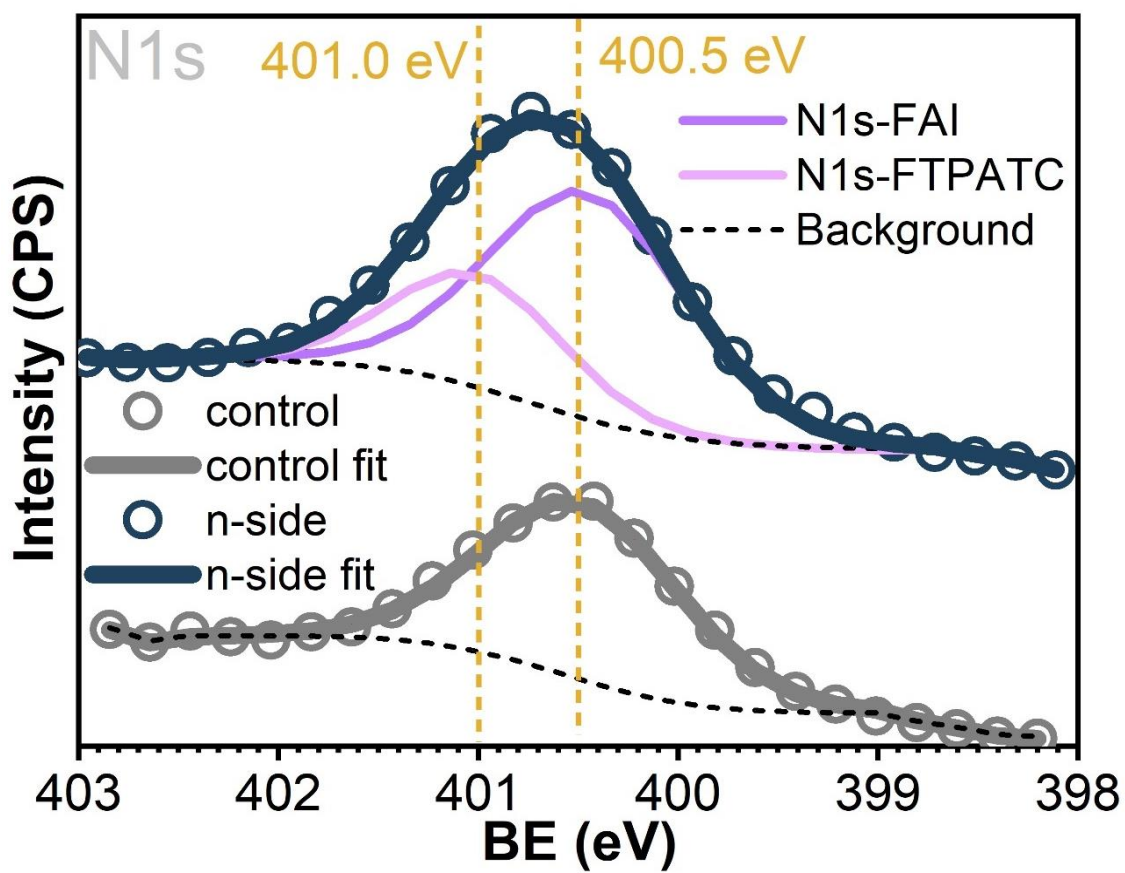

Figure S17 – N1s spectra of control and n-side perovskite samples

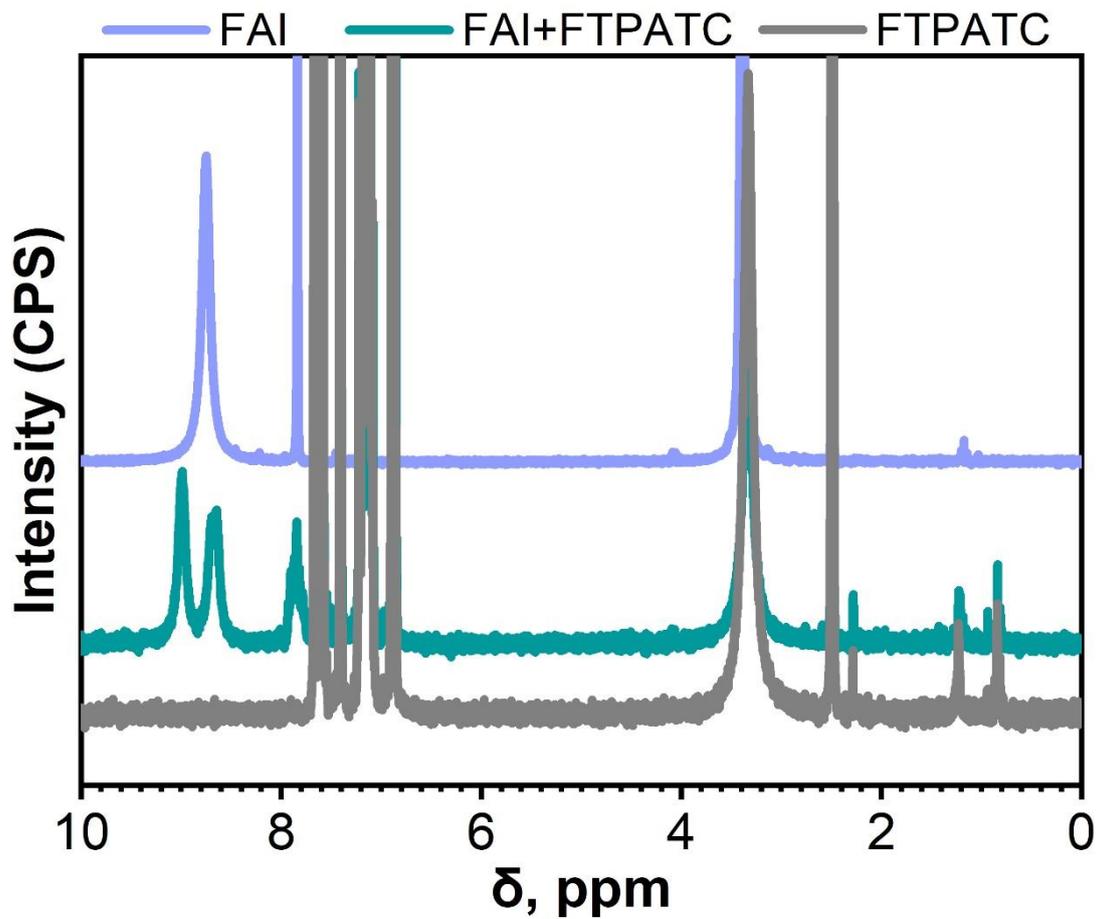

Figure S18 – $^1$H NMR spectrum of FTPATC, FAI and its mix in DMSO solution

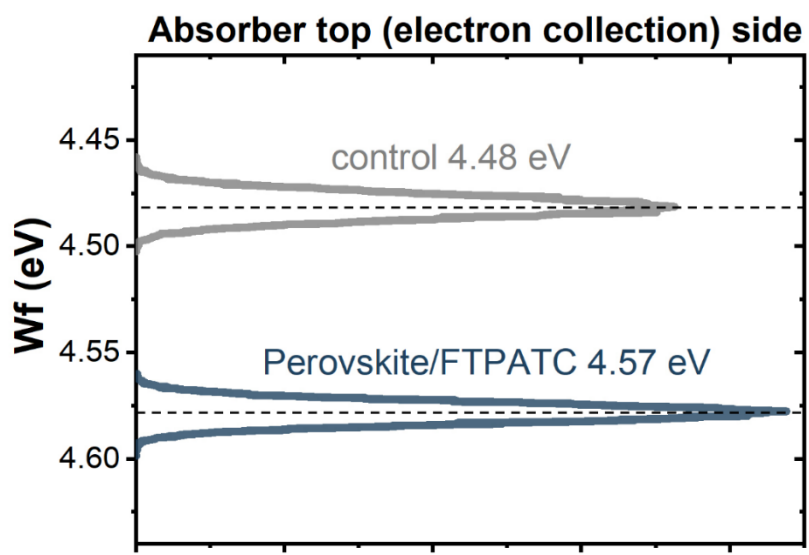

Figure S19 – The $W_f$ values measured for the top (electron) collection side of the bare perovskite absorber and perovskite absorber with FTPATC interlayer

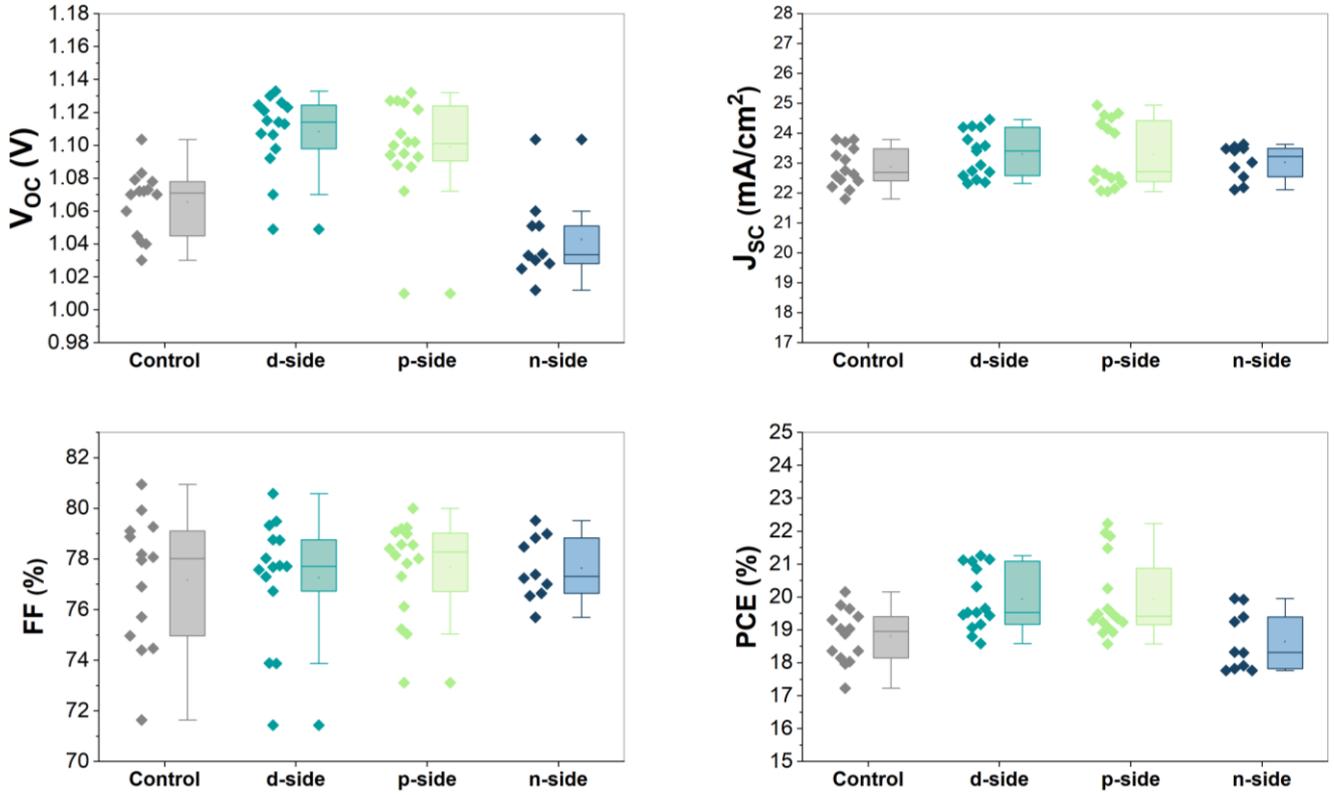

Figure S20 – The box-charts with output data for the fabricated PSCs - $V_{oc}$ (a), Jsc (b), FF (c), and PCE (d)

**Details for the double diode model used for fitting of dark JV curves**

The equation for the diode current-voltage curve, including series (Rs) and shunt (Rsh) resistance, has the following expression (S1):

$$J = J_0 \cdot \left( \exp\left( \frac{q(V - J \cdot R_s)}{m \cdot k \cdot T} \right) - 1 \right) + \frac{V - J \cdot R_s}{R_{sh}} . \quad (S1)$$

For common pn junction, such a model describes the characteristics of real structures quite accurately. However, for a p-i-n structure, such a model does not always allow us to express the characteristic corresponding to experimental results. The reason for this is the presence of two barriers from the p- and n-regions, therefore, to calculate the current-voltage characteristics of the p-i-n structure, we used a two-diode model, represented by the equivalent circuit in Fig. S2.[1]

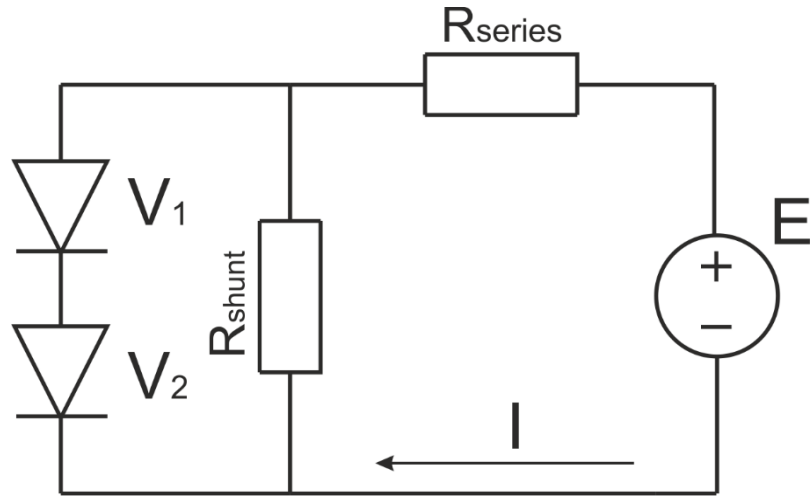

Figure S21 – Double diode circuit for modeling of pin PSC

Diodes 1 and 2 in this circuit are ideal diodes, the current-voltage characteristics of which are described by the expressions S2-S3:

$$J_1 = J_{01} \cdot \left( \exp\left( \frac{qV_1}{m_1 \cdot k \cdot T} \right) - 1 \right) \tag{S2}$$

$$J_2 = J_{02} \cdot \left( \exp\left( \frac{qV_2}{m_2 \cdot k \cdot T} \right) - 1 \right) \tag{S3}$$

Two diode structures are connected in series, so the currents are equal to each other.

$$J_d = J_{01} \cdot \left( \exp\left( \frac{qV_1}{m_1 \cdot k \cdot T} \right) - 1 \right) = J_{02} \cdot \left( \exp\left( \frac{qV_2}{m_2 \cdot k \cdot T} \right) - 1 \right) \tag{S4}$$

Since the equivalent circuit is branched, it is necessary to solve a system of equations obtained from Kirchhoff's laws to calculate the current-voltage characteristic, For a given voltage V, unknown values are the currents in the diode and shunt resistance circuits $J_d$ and Rsh. The voltages applied to each diode $V_1$ and V2, and the total current J, which we must find for each given voltage. The system of equations (S4) - (S7) is sufficient for the numerical calculation of the current – voltage characteristics according to the two-diode model, but it is necessary to calculate the model parameters: leakage currents of each individual diode $J_{01}$ and $J_{02}$, non-ideality coefficients $m_1$ and $m_2$ of each diode, series resistance $R_s$ and shunt resistance $R_{sh}$.

$$V_1 + V_2 = V - J \cdot R_s, \tag{S5}$$

$$J_{R_{sh}} = \frac{V - J \cdot R_s}{R_{sh}}, \tag{S6}$$

$$J = J_d + J_{R_{sh}}. \tag{S7}$$

The calculation was done with one of the methods for multi-parameter optimization, in which the objective function requiring minimization of the sum for the differences between the experimental and theoretically calculated currents (S8).

$$\sum_{i=1}^{m}(J_{ex_i} - J_{teor_i})^2, \tag{S8}$$

where m is the number of experimental points.

Calculation program was developed in Borland Delphi 7, which allows determining the parameters of a two-diode structure using multi-parameter optimization by the coordinate descent method.

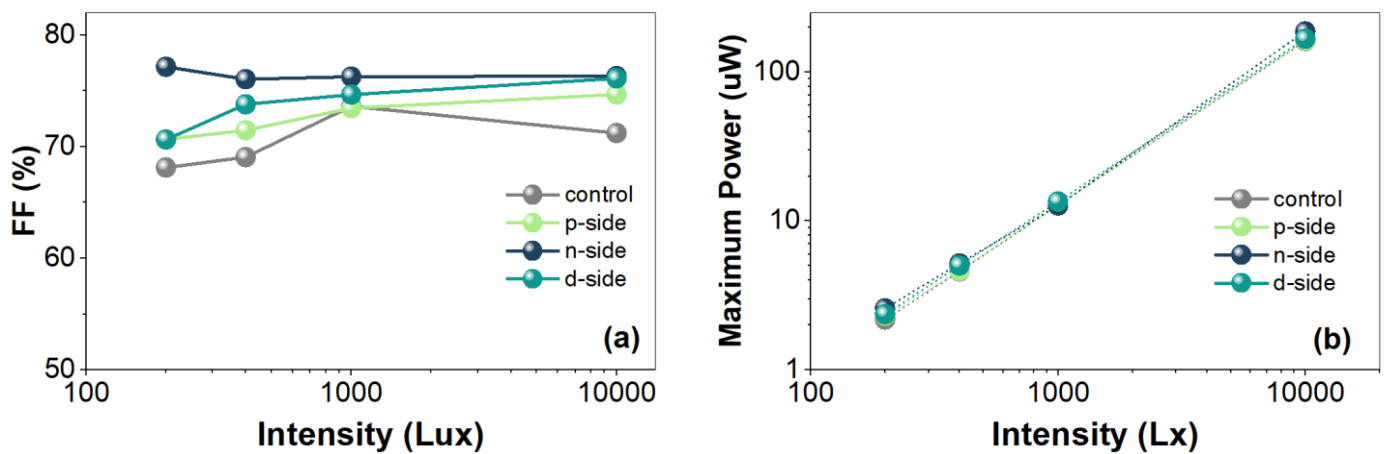

Figure S22 – Intensity-dependent perfromance of FF vs $I_0$ (a);
Intensity-dependent perfromance of $P_{max}$ vs $I_0$ (b)